\documentclass[twocolumn]{IEEEtran}
\usepackage{amsthm,amssymb,graphicx,multirow,amsmath,color,amsfonts}
\usepackage[update,prepend]{epstopdf}
\usepackage[noadjust]{cite}
\usepackage[latin1]{inputenc}
\usepackage{tikz}
\usetikzlibrary{arrows,calc}		
\usepackage{bbm} 
\usepackage{pdfpages}
\usepackage{tabulary}
\usepackage{multirow}
\usepackage{subfigure}

\include{notation}

\usepackage{comment}

\newcommand*{\Scale}[2][4]{\scalebox{#1}{$#2$}}%

\def\dar2add#1{{\color{green}#1}}

\def \calCfunc {\mathcal{C}}

\allowdisplaybreaks 

\begin{document}
\graphicspath{{./Figures/}}

\newtheorem{lemma}{Lemma}
\newtheorem{corollary}{Corollary}
\newtheorem{theorem}{Theorem}
\newtheorem{remark}{Remark}
\newtheorem{example}{Example}
\newtheorem{definition}{Definition}

\title{Downlink Multi-Antenna Heterogeneous Cellular Network with Load Balancing
}
\author{Abhishek K. Gupta, Harpreet S. Dhillon,~\IEEEmembership{Member, IEEE}, Sriram Vishwanath,~\IEEEmembership{Senior Member, IEEE}, \\Jeffrey G. Andrews,~\IEEEmembership{Fellow, IEEE}
\thanks{Manuscript received October 24, 2013; revised April 4, 2014 and July 17, 2014; accepted August 10, 2014. The associate editor coordinating the review of this paper and approving it for
publication was O. Oyman.}
\thanks{A. K. Gupta, S. Vishwanath and J. G. Andrews are with the Wireless Networking and Communications Group (WNCG), The University of Texas at Austin, TX (email: g.kr.abhishek@utexas.edu, sriram@ece.utexas.edu and jandrews@ece.utexas.edu).} 
\thanks{H. S. Dhillon is with the Wireless@VT, Department of Electrical and Computer Engineering, Virginia Tech, Blacksburg, VA (email: hdhillon@vt.edu).}
\thanks{This paper will be presented in part at the IEEE Globecom, Austin, TX, Dec. 2014~\cite{GupDhiC2014}. \hfill Last revised: \today.
}
}

\maketitle

\begin{abstract}
We model and analyze heterogeneous cellular networks with multiple antenna BSs (multi-antenna HetNets) with $K$ classes or tiers of base stations (BSs), which may differ in terms of transmit power, deployment density, number of transmit antennas, number of users served, transmission scheme, and path loss exponent. We show that the cell selection rules in multi-antenna HetNets may differ significantly from the single-antenna HetNets due to the possible differences in multi-antenna transmission schemes across tiers. While it is challenging to derive exact cell selection rules even for maximizing signal-to-interference-plus-noise-ratio (SINR) at the receiver, we show that adding an appropriately chosen tier-dependent cell selection bias in the received power yields a close approximation. Assuming arbitrary selection bias for each tier, simple expressions for downlink coverage and rate are derived. For coverage maximization, the required selection bias for each tier is given in closed form. Due to this connection with biasing, multi-antenna HetNets may balance load more naturally across tiers in certain regimes compared to single-antenna HetNets, where a large cell selection bias is often needed to offload traffic to small cells. 
\end{abstract}

\begin{keywords}
Multi-antenna heterogeneous cellular network, stochastic geometry, coverage probability, cell selection bias.
\end{keywords}

\section{Introduction}
Current cellular networks are undergoing a significant transformation due to the capacity-driven opportunistic deployment of small cells~\cite{AndJ2013}. The resulting cell splitting gain due to increased infrastructure, along with the possibility of multi-antenna transmission, provides one of the most promising solutions to handle current data deluge~\cite{CisM2012}. Due to the relative maturity of both HetNets~\cite{QuaM2011} and multi-antenna transmission techniques in cellular standards~\cite{Li2010}, their coexistence in future networks is almost inevitable. However, quite remarkably, this synergy is not reflected in the current HetNet research efforts using tools from stochastic geometry, e.g., see~\cite{ElSHosJ2013} for a survey. 
These two strategies are mostly studied in isolation, thereby missing important interplay between them, especially in the cases where one can complement the other's shortcomings. In this paper, we take a step towards bridging this gap. In particular, we focus on developing simple yet useful cell selection rules for multi-antenna HetNets, which highlight certain natural connections with current standardization activities, most importantly cell range expansion and load balancing~\cite{DamMonJ2011}.

\subsection{Related Work}
The increasing heterogeneity in cellular infrastructure and irregularity in BS locations, have driven popular cellular models, such as deterministic grid and Wyner models towards obsolescence~\cite{AndBacJ2011}. A natural way to model the inherent randomness in such networks is by using random spatial models. Modeling the BS locations as a Poisson Point Process (PPP) further lends tractability and allows to compute simple expressions for key metrics, such as coverage and rate. This model was first introduced for multi-tier cellular networks in~\cite{DhiGanC2011,DhiGanJ2012}, and generalized further in~\cite{JoSanJ2012,MadResJ2012a,MukJ2012,DhiAndJ2014} to study various cell selection rules for single-antenna HetNets. Interested readers can refer to~\cite{ElSHosJ2013} for a detailed survey and to~\cite{Hanley,Yates,HongLuo,Kuang} for a subset of prior work on cell selection. The PPP-based HetNet model of~\cite{DhiGanJ2012} was generalized to multi-antenna HetNets in~\cite{DhiKouC2012}, where an upper bound on downlink coverage probability was derived using tools from stochastic geometry, and in~\cite{DhiKouJ2013}, where stochastic orders were used to order the performance of various multi-antenna transmission schemes in a general multi-antenna HetNet. A key difference between these works and the current paper is that they all assume maximum instantaneous SINR based cell selection, whereas this paper assumes average-SINR based cell selection with an additional flexibility of per-tier cell selection bias.
 Another relevant prior work is~\cite{HeaKouJ2013}, which studies a hybrid multi-antenna HetNet with a fixed size ``typical'' cell. In this paper, we extend the multi-antenna HetNet model of~\cite{DhiKouC2012,DhiKouJ2013}, with a special emphasis on cell selection and its impact on downlink coverage probability and average rate. 

It should be noted that while there is a limited prior work on general multi-antenna HetNets besides~\cite{DhiKouC2012,DhiKouJ2013,HeaKouJ2013}, the special case of two-tier multi-antenna HetNets has been investigated fairly thoroughly. For example,~\cite{ChaKouAnd2009} compares single and multi-user linear beamforming in a two-tier network under perfect channel state information (CSI), \cite{Park2010}, \cite{Park2011}, respectively study random orthogonal beamforming with max-rate scheduling and coordinated beamforming for two-tier networks, and~\cite{AkoKouHea11} studies the effect of channel uncertainty in linear beamforming. Additionally, it is worth mentioning that multi-antenna transmission schemes have been investigated extensively in the context of ad-hoc networks, see e.g.,~\cite{HunWebAnd08, LouKayCol11,VazHea12}. Some of the tools developed there can be extended to HetNets.

\subsection{Contributions}


{\em Cell selection rules for multi-antenna HetNets.} We investigate cell selection rules for a multi-antenna HetNet consisting of $K$ different classes of BSs, which may differ in terms of transmit power, deployment density, number of transmit antennas, number of users served, transmission scheme, and path loss exponent. This is motivated by the fact that, unlike single-antenna HetNet, connecting with the BS that provides maximum received power may not always maximize the received SINR in a multi-antenna HetNet. Building upon this observation, we show that although it is challenging to derive exact cell selection rule to maximize average received SINR, a simpler selection rule based on adding an appropriately chosen per-tier selection bias in the received power yields a surprisingly close approximation. We also derive this approximate per-tier selection bias in closed form for coverage probability maximization. One key observation is that the bias value depends only on the number of antennas at the BS and the number of users served in each resource block, which makes it easier to implement it in practice. Assuming a general cell selection bias for each tier, we derive exact expressions for downlink SINR distribution, from which we study downlink rate achievable at a typical user.


{\em Connections with biasing and load balancing.} An important interpretation of our results is in terms of {\em load balancing} in HetNets. It is well known that the load in HetNets is unbalanced due to differences in transmit powers of macro and small cells~\cite{DhiGanJ2013}. An artificial bias is generally introduced to expand the coverage regions of small cells in order to {\em offload} more users from macrocells~\cite{SinDhiJ2013,Ye,AndSinJ2014}. However, as discussed above, multi-antenna HetNets may need selection bias even for SINR, and hence coverage, maximization. Therefore, in certain cases, the selection bias that maximizes coverage, may naturally balance load across tiers compared to single-antenna HetNets. We characterize such regimes in this paper and validate our intuition through extensive simulations.

\begin{figure}
\centering
\begin{center}
\includegraphics[width=.7\columnwidth,trim=10 40 40 70,clip=true]{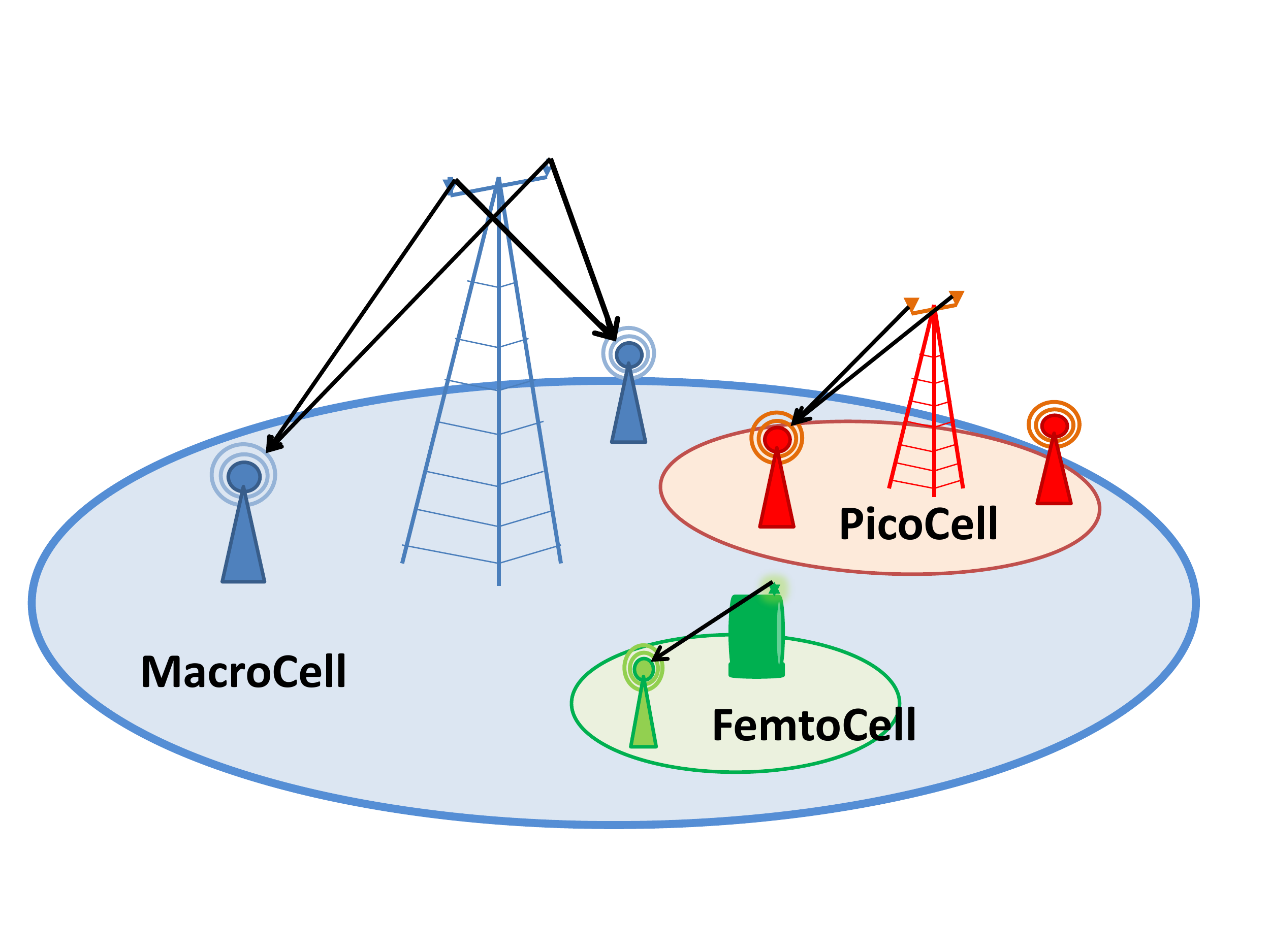}
\caption{A three-tier HetNet consisting of macro, pico and femtocells, with different number of transmit antennas and different transmission schemes across tiers.}
\label{fig:HetNet}
\end{center}
\end{figure}

\section{System Model} \label{sec:SysMod}


We consider a $K$-tier HetNet consisting of $K$ different tiers or classes of BSs. For notational ease we denote $\mathcal{K} = \{1, 2, \ldots K\}$. The BSs across tiers differ in terms of the transmit power $P_k$ (to each user), deployment density $\lambda_k$, number of transmit antennas $M_k$, number of users served in each resource block $\Psi_k$, transmission scheme, and the path-loss exponent $\alpha_k$. The locations of each tier are assumed to be sampled from an independent homogeneous PPP $\Phi_k$ of density $\lambda_k$. This model is the same as the one introduced in~\cite{DhiKouJ2013}, except for some key differences in cell selection, which will be clarified later in this section. Although the PPP model is likely more accurate for the opportunistic deployment of small cells, it has also been verified for the planned tiers, such as macrocells, both by empirical evidence~\cite{TayDhiC2012} and theoretical validation~\cite{BlaKarJ2012} under sufficient channel randomness. For easier exposition later in the paper, we denote the locations of all the BSs by $\Phi = \cup_{k \in \mathcal{K}} \Phi_k$. A particular realization of $\Phi$ will be denoted by $\phi$. Each user is assumed to have a single receive antenna. Note that since each BS serves multiple users and the channel from a BS to each user is multiple-input single-output (MISO), the current setup can be precisely defined as a $K$-tier multi-user MISO HetNet. The temporal evolution of the system is not studied in this paper.

\begin{table}
\caption{Notations Summary}
\begin{tabulary}{\columnwidth}{ |c | C | }\hline
{\bf Notation} &{\bf Description}\\ \hline
$\mathcal{K}$ & Indices of the tiers, where $\mathcal{K} = \{1,2,\ldots, K\}$ \\ \hline
$\Phi_k, \Phi; \phi$ & A PPP modeling the locations of $k^{th}$ tier BSs, $\Phi = \cup_{k \in \mathcal{K}} \Phi_k$; a realization of $\Phi$\\ \hline
$\Phi_u$ & An independent PPP modeling user locations\\ \hline
$\mathbbm{1}(e),\mathbbm{1}_e$ & Indicator function for logic $e$\\ \hline
$P_k,\lambda_k, \alpha_k$  & Downlink transmit power to each user, deployment
density, path loss exponent of the $k^{th}$ tier BSs\\ \hline
$M_k,\Psi_k,\Delta_k$ & Number of transmit antennas, number of users served in
each resource block by a $k^{th}$ tier BS, $\Delta_k=M_k-\Psi_k+1$\\ \hline
$h_{x_kk}$ & Channel power distribution of the direct link from a BS at $x_k \in \Phi_k$ to a typical user, $h_{x_kk} \sim \Gamma(\Delta_k; 1)$\\ \hline
$g_{y_jj}$ & Channel power distribution of the interference link from a BS $y_j \in \Phi_j$ to a typical user, $g_{y_jj}\sim \Gamma(\Psi_j; 1)$\\ \hline
$\widehat{{(\cdot)}}_j$ & Ratio of a $j^{th}$ parameter to the same parameter of the serving tier, e.g., if  $k^{th}$ tier is serving, $\widehat{P}_j=\frac{P_j}{P_k}$ \\ \hline
$B_k,A_k$& Cell selection bias, selection probability for $k^{th}$ tier\\ \hline
$\gamma_k,P_c$ & Instantaneous SINR, coverage probability\\ \hline
$R_k, R_c$ & Instantaneous rate conditional on serving BS being in $k^{th}$ tier, rate coverage \\ \hline
$W_k,\mathcal{O}_k,N_k$ &  Total time frequency resource, e.g., bandwidth, for each $k^{th}$ tier BS, fraction of resources allocated to each user served by $k^{th}$ tier, average load over a $k^{th}$ tier BS\\ \hline
\end{tabulary}
\end{table}

For multiple access, we assume orthogonal resource partitioning, e.g., orthogonal frequency division multiple access (OFDMA), with a provision that multiple users can be scheduled on a given resource block if the BS has sufficient degrees of freedom to orthogonalize them in space. In terms of the notation introduced above, this leads to the constraint $\Psi_k \leq M_k$ for all $k \in \mathcal{K}$. The users are assumed to form an independent PPP $\Phi_u$ of density $\lambda_u$. Note that more sophisticated user location models can in principle be considered, e.g., by using tools from~\cite{DhiGanJ2013b}, but are out of the scope of this paper. The downlink analysis will be performed at a typical user, which is assumed to be at the origin. This is facilitated by Slivnyak's theorem, which states that the properties observed by a typical point of the point process $\Phi_u$ are the same as those observed by the origin in the point process $\Phi_u \cup \{0\}$~\cite{StoKenB1995}.

In this paper, we will restrict our discussion to zero-forcing precoding, which is general enough to encompass important transmission schemes such as beamforming and spatial division multiple access (SDMA), while being tractable enough to provide important system design guidelines. Note that due to precoding at the BS, the effective channel gain to a given user depends upon whether that BS acts as a serving BS or an interferer for that user. For example, if a multi-antenna transmitter beamforms to a given user, the effective channel gain would be much higher than when it simply acts as an interferer. Therefore, we denote the effective channel power gain from a $k^{th}$ tier BS located at $x_k \in \Phi_k$ to a typical user by $h_{x_k k}$ when it acts as a serving BS, and by $g_{x_k k}$ when it acts as an interferer. We also assume perfect CSI at the transmitter, although as argued in~\cite{DhiKouJ2013} the tools developed in this paper can also be used to study the effect of imperfect CSI on the network performance. For zero-forcing precoding with perfect CSI under Rayleigh fading, $h_{x_k k} \sim \Gamma(\Delta_k, 1)$ and $g_{x_k k} \sim \Gamma(\Psi_k, 1)$, where $\Delta_i= M_i-\Psi_i+1$ as discussed in detail in \cite{DhiKouJ2013}. More general precoding techniques are left for future work. The received power at a typical user from a serving BS located at $x_k \in \Phi_k$ is
\begin{align}
P(x_k) = P_k h_{x_k k} \|x_k\|^{-\alpha_k},
\end{align}
where $\|x_k\|^{-\alpha_k}$ is a standard power-law path-loss with exponent $\alpha_k$, which may be different for different tiers. Also recall that $P_k$ is the transmit power to each user. The resulting SINR $\gamma_k(x_k)$ is
\begin{align}
\gamma_k(x_k)=\frac{P_kh_{x_kk} {\|x_k\|}^{-\alpha_k}}{I+N},
\label{expSINR}
\end{align}
where $N$ is the noise power, and $I$ is the interference power given by
\begin{align}
I=\sum_{j\in\mathcal{K}}{I_j} =\sum_{j\in\mathcal{K}}{ \sum_{x \in \Phi_j\backslash x_k} {P_j g_{xj} {\|x\|}^{-\alpha_j}}}.
\end{align}
For cleaner exposition in the next section, we denote the average received power from a $k^{th}$ tier BS by $P_{rk} (x_k)$, which can be expressed as
\begin{align}
P_{rk}(x_k) = P_k \Delta_k \|x_k\|^{-\alpha_k}.
\end{align}

For this setup, we  discuss cell selection for multi-antenna HetNets in the next section. As evident from the following discussion, the cell selection principles for multi-antenna HetNets have some fundamental differences compared to their counterparts in single-input single-output (SISO) HetNets, mainly because of the precoding at the transmitter.

\section{Cell Selection} \label{sec:cell_selection}
Recall that a usual cell selection criterion in SISO HetNets is to connect to the BS that provides the maximum average received power, possibly with a certain bias value for load balancing~\cite{AndSinJ2014}. In the case when there is no bias, this cell selection criterion is also the one that maximizes the SINR. Therefore, to maximize coverage probability, a typical user simply connects to the BS that provides the highest received power, as discussed in~\cite{JoSanJ2012}. However, it is easy to construct a simple toy example showing that this is not the case in multi-antenna HetNets.

\begin{example}
 Consider two BSs at the same distance from a typical user, one having $4$ antennas serving a single user, which with a slight overloading of the notation implies $M_1=4, \Psi_1 = 1, \Delta_1 = 4$, and the other serving $4$ users with $6$ antennas, i.e., $M_2 = 6, \Psi_2 = 4, \Delta_2 = 3$. Since $\Delta_1 > \Delta_2$, a typical user should be served by the first BS to maximize average received power. However, since $\Psi_2>\Psi_1$, the interference from second BS is larger than the first and it should be served by the second to minimize the received interference power. Further, since $\frac{\Delta_2}{\Psi_1} > \frac{\Delta_1}{\Psi_2}$, it should be served by the second BS to maximize its received SINR.
\end{example}

We first discuss cell selection with the goal of maximizing the average received SINR conditional on the point process $\Phi$. Note that to maximize both the average received power and average SINR, it is strictly suboptimal for a typical user to connect to any BS except the ones that are closest to it in each tier. We denote by $\gamma_k$ the instantaneous SINR when a typical user connects to the closest $k^{th}$ tier BS. Under maximum average SINR cell selection rule, $k^{th}$ tier is selected if
\begin{align}
k = \arg \max_{j \in \mathcal{K}}\mathbb{E}[\gamma_j | \Phi = \phi],
\label{eq:meanSINR_eq}
\end{align}
where the selection rule clearly depends upon the distances of the BSs to a typical mobile due to conditioning on the realization of the point process. As will be evident in the sequel, it is quite challenging to derive an exact selection rule from~\eqref{eq:meanSINR_eq}. We take several alternate routes in the following subsections. Our eventual goal is to come up a simple and practical cell selection rule that works well across wide range of system parameters.

\subsection{Results from Stochastic Orders} \label{sec:stocdom}
Put simply, stochastic orders are binary relations defined to compare random variables, see~\cite{ShaShaB2007} for details. 
Ideas from stochastic orders were used in~\cite{DhiKouJ2013} to compare the coverage probability and downlink rate for various multi-antenna transmission schemes. Using tools developed in~\cite{DhiKouJ2013}, it is possible to derive a sufficient condition for the selection rule that is slightly stronger than the average SINR maximization. The idea is to condition on the locations of the BSs and then find a condition under which the conditional cumulative distribution function (CCDF) of SINR from the chosen tier dominates the conditional CCDF of SINR for all other choices. More formally, the $k^{th}$ tier is selected if
\begin{align}
k = \arg \max_{j \in \mathcal{K}} \mathbb{P}(\gamma_j > z | \Phi = \phi), \forall z.
\label{eq:stocdom}
\end{align}
In stochastic ordering terms, this means that conditional on the point process $\Phi$, $\gamma_k$ (first order) stochastically dominates $\gamma_j$ for all $j \in \mathcal{K}$, which is denoted by $\gamma_k {\ge}_{st} \gamma_j \forall j \in \mathcal{K} \setminus \{k\}$. Clearly, $\eqref{eq:stocdom} \Rightarrow \eqref{eq:meanSINR_eq}$. Now let us rewrite $\gamma_k$ and $\gamma_j$ as
\begin{align}
\gamma_k(x_k)&=\frac{P_kh_{x_kk} {\|x_k\|}^{-\alpha_k}}{P_jg_{x_jj} {\|x_j\|}^{-\alpha_j}+W} =\frac{h_{x_kk}a_k}{g_{x_jj}a_j+W} \label{eq:7}\\
\gamma_j(x_j)&=\frac{P_j h_{x_j j} {\|x_j\|}^{-\alpha_j}}{P_k g_{x_k k} {\|x_k\|}^{-\alpha_k}+W} =\frac{h_{x_j j}a_j}{g_{x_k k}a_k+W} \label{eq:8},
\end{align}
where $a_i=P_i {\|x_i\|}^{-\alpha_i}$ and $W$ is  a random variable representing thermal noise plus interference from all the BSs except $x_k \in \Phi_k$ and $x_j \in \Phi_j$. For $a_k\ge a_j$, using~\cite[Lemma 3]{DhiKouJ2013}, it can be shown that conditional on the locations of the BSs, $\gamma_k {\ge}_{st} \gamma_j, \forall j \in \mathcal{K},$ if $\Delta_k\ge \Delta_j$ and $\Psi_k\ge \Psi_j$.
This leads to the following set of sufficient conditions for \eqref{eq:stocdom} to hold:
\begin{align}
P_k\|x_k\|^{-\alpha_k} &\ge P_i\|x_i\|^{-\alpha_i} \ \ \forall i \ne k \label{Eq:SO11}\\
\Delta_k &\ge \Delta_i\ \ \forall i \ne k\label{Eq:SO12}\\
\Psi_k &\ge \Psi_i\ \ \forall i \ne k.
\label{Eq:SO13}
\end{align}
Clearly, these conditions are also sufficient for the selection of $k^{th}$ tier to maximize the average received SINR. Since first order stochastic dominance is a stronger notion than the ordering of the means required in~\eqref{eq:meanSINR_eq}, it is possible that none of the tiers satisfy the above set of conditions simultaneously, which limits the applicability of this selection rule. In the pursuit of a more useful cell selection rule, we explore two more directions below.

\subsection{Results from Jenson's Inequality Approximation}
In this section, we approximate the mean SINR using Jenson's inequality. For any realization of point process $\phi$, the mean of SINR over fading distribution is given by 
\begin{align}
\mathbb{E}[\gamma_k(x_k)| \phi]&=\mathbb{E}\left[\frac{h_{x_kk}a_k}{g_{x_jj}a_j+W}| \phi\right] \\
&=a_k \Delta_k \mathbb{E}\left[\frac{1}{g_{x_jj}a_j+W}| \phi\right] \\
&\ge  a_k \Delta_k \frac{1}{\mathbb{E}[g_{x_jj}a_j+W | \phi]}=\frac{a_k \Delta_k} {\Psi_j a_j+\mathbb{E}[W| \phi]}.
\end{align}
Using this lower bound, the conditions for~\eqref{eq:meanSINR_eq} to hold are 
\begin{align}
P_k\Delta_k\|x_k\|^{-\alpha_k} &\ge P_i\Delta_i\|x_i\|^{-\alpha_i} \ \ \forall i \ne k \label{Eq:SO1}\\
P_k\Psi_k\|x_k\|^{-\alpha_k} &\ge P_i\Psi_i\|x_i\|^{-\alpha_i} \ \ \forall i \ne k.
\label{Eq:SO2}
\end{align}
Although this approximation has reduced the number of simultaneous conditions from three to two compared to the previous subsection, it is still possible that none of the tiers satisfy these new conditions simultaneously. We construct an example below to highlight this point. The same example is also true for the conditions \eqref{Eq:SO11} \eqref{Eq:SO12} and \eqref{Eq:SO13} discussed in the previous subsection in the context of stochastic orders.

\begin{example}
Consider a two-tier network such that the distance of the typical user to the nearest BS in each tier is the same, i.e., $\|x_1\|=\|x_2\|$. Further assume that the transmit powers for the two tiers are also the same. Fixing $\Delta_1=2, \Delta_2=1, \Psi_1=2, \Psi_2=3$, it is easy to check that neither $k=1$ nor $k=2$ satisfy \eqref{Eq:SO1} and \eqref{Eq:SO2} simultaneously.
\end{example}

As discussed above, although there is no guarantee that the conditions given by \eqref{Eq:SO1} and \eqref{Eq:SO2} would provide a solution, it is possible to derive a simpler but more useful condition for the selection of $k^{th}$ tier by combining \eqref{Eq:SO1} and \eqref{Eq:SO2}. The new selection law is
\begin{align}
P_k\sqrt{\Delta_k\Psi_k}\|x_k\|^{-\alpha_k} \ge P_i\sqrt{\Delta_i\Psi_i}\|x_i\|^{-\alpha_i} \ \ \forall i \ne k. \label{Eq:SO3} \end{align}
It is easy to verify that there always exists a $k \in \mathcal{K}$ for which the selection law \eqref{Eq:SO3} holds and is equal to solution of \eqref{Eq:SO1} and \eqref{Eq:SO2} if the solution of latter pair exists. We now remark on an interesting connection between this cell selection criterion and the idea of cell selection bias used for load balancing.

\begin{remark}[Connections with biasing] \label{rem:bias_connection}
Note that~\eqref{Eq:SO3} can be equivalently expressed as:
\begin{align}
\sqrt{\frac{\Psi_k}{\Delta_k}}P_k \Delta_k {\|x_k\|}^{-\alpha_k} \ge \sqrt{\frac{\Psi_i}{\Delta_i}} P_i \Delta_i {\|x_i\|}^{-\alpha_i} \ \ \forall i \ne k,
\label{Eq:SO3_eq}
\end{align}
where recall that $P_{rk} (x_k) = P_k \Delta_k {\|x_k\|}^{-\alpha_k}$ is the average received power from a $k^{th}$ tier BS located at $x_k \in \Phi_k$. The selection criterion can now be expressed in terms of the average received power as:
\begin{align}
\sqrt{\frac{\Psi_k}{\Delta_k}} P_{rk} (x_k) \geq  \sqrt{\frac{\Psi_i}{\Delta_i}} P_{ri} (x_i)  \ \ \forall i \ne k,\label{eq:biaspsibydelta}
\end{align}
where $B_k = \sqrt{\frac{\Psi_k}{\Delta_k}}$ can be perceived as a ``cell selection bias''. Note that since $\Psi_k = \Delta_k = 1$ $\forall k \in \mathcal{K}$ in the SISO case, the above rule reduces to selecting the BS providing highest received power, which is also the SINR maximizing rule in the SISO case.
\end{remark}

Although \eqref{Eq:SO3} will be shown to be a more useful cell selection rule, it is in principle possible to derive other candidate laws, e.g., by adding the inequalities \eqref{Eq:SO1} and \eqref{Eq:SO2}, we get:
\begin{align}
P_k\left(\Delta_k +\Psi_k\right){\|x_k\|}^{-\alpha_k} \ge P_i\left(\Delta_i+\Psi_i\right) {\|x_i\|}^{-\alpha_i} \ \ \forall i \ne k, \label{Eq:SO4}
\end{align}
which is also consistent with the biasing interpretation discussed in Remark~\ref{rem:bias_connection}. The bias that needs to be added to the received power (in dB) in this case is $B_k = 1+ \frac{\Psi_k}{\Delta_k}$. We will comment more on this alternate cell selection law in the numerical results section. After gaining these insights, we now investigate the cell selection criterion using the mean SINR expression in the following subsection.

\subsection{Results from Mean SINR Expression}
We first present an exact expression for the mean SINR. Using this expression, we will argue that the general form of the cell selection criterion given by~\eqref{Eq:SO3} is possibly the more appropriate one to maximize the average SINR, which will further be validated in the numerical results section.

\begin{lemma}\label{lemma:meansinr}
Given a realization $\phi$ of point process $\Phi$, mean SINR at typical user (at origin)  associated with a $k^{th}$ tier BS situated at $x_k$ is given as
\begin{align}
\mathbb{E}[\gamma_k|\phi]= \frac{P_k\Delta_k}{ {\|x_k\|}^{\alpha_k}}\int_0^{\infty}{
    \frac{e^{-NT}}
    {\prod_{y\in \phi \backslash {x_k}}{
        {(1+\frac{TP_j}{{\|y\|}^{\alpha_j}})}^{\Psi_j}
        }        }{\rm d}T.
    }\label{eq:meansinrdef}
\end{align}
\end{lemma}
\begin{IEEEproof}
\begin{eqnarray*}
\mathbb{E}[\gamma_k|\phi]&=& \mathbb{E}\left[\frac{P_kh_{x_kk}{\|x_k\|}^{-\alpha_k}}
                        {N+\sum_{j\in\mathcal{K}}\sum_{y\in \phi_j\backslash {x_k}}{P_jg_{yj}{\|y\|}^{-\alpha_j}}}\right]\\
                        &=&\frac{P_k\mathbb{E}[h_{x_kk}]}{{\|x_k\|}^{\alpha_k}}\mathbb{E}\left[\frac{1}
                        {N+\sum_{j\in\mathcal{K}}\sum_{y\in \phi_j\backslash {x_k}}\frac{P_jg_{yj}}{{\|y\|}^{\alpha_j}}}\right]
                        \end{eqnarray*}
Let $F$ be an independent exponential random variable with mean $1$, then the last term in the expression can be written as
\begin{eqnarray*}
\mathbb{E}\left[\frac{\mathbb{E}_F[{F}]}
{N+\sum_{j\in\mathcal{K}}\sum_{y\in \phi_j\backslash {x_k}}\frac{P_jg_y}{{\|y\|}^{\alpha_j}}}\right]&\ &\ \ \ \ \ \
\end{eqnarray*}
\begin{eqnarray*}
&=&\mathbb{E}\left[\mathbb{E}_F\left[\frac{F}
{N+\sum_{j\in\mathcal{K}}\sum_{y\in \phi_j\backslash {x_k}}\frac{P_jg_y}{{\|y\|}^{\alpha_j}}}\right]\right]\\
                        &=&\int_0^{\infty}
                        {\mathbb{P}\left[\frac{F}
                        {N+\sum_{j\in\mathcal{K}}\sum_{y\in \phi_j\backslash {x_k}}\frac{P_jg_{yj}}{{\|y\|}^{\alpha_j}}}>T\right] {\rm d}T}\\
                        &=&\int_0^{\infty}
                        {\mathbb{E}[e^{-NT-T\sum_{j\in\mathcal{K}}\sum_{y\in \phi_j \backslash {x_k}}{P_jg_{{yj}}{\|y\|}^{-\alpha_j}}} ]{\rm d}T}\\
                        &=&\int_0^{\infty}
                        {e^{-NT}\prod_{j\in\mathcal{K}}\prod_{y\in \phi_j \backslash 
                        {x_k}}\mathbb{E}[e^{-T{P_jg_{{yj}}{\|y\|}^{-\alpha_j}}} ]{\rm d}T}\\
                        &=&\int_0^{\infty}
                        {e^{-NT}\prod_{j\in\mathcal{K}}\prod_{y\in \phi_j \backslash {x_k}}\frac{1}{{(1+{T{P_j}{{\|y\|}^{-\alpha_j}}})}^{\Psi_j}}{\rm d}T}.
                        \end{eqnarray*}
                        Note that the insertion of random variable $F$ makes the calculation simple and does not change the integral due to independence assumption.
\end{IEEEproof}

Using the same setup as Section~\ref{sec:stocdom}, where we focused on the selection of two BSs $x_k \in \Phi_k$ and $x_j \in \Phi_j$, and represented to thermal noise plus interference from all the other BSs except $x_k$ and $x_j$ by $W$, we can argue that a simple selection criterion very similar to the one given by~\eqref{Eq:SO3} can be derived directly form the expressions of the mean SINR. The result is given in the following Lemma.

\begin{lemma}\label{lemma:meansinrw0} The mean SINR of a $k^{th}$ tier BS at $x_k$ at typical user at origin is greater than the mean SINR of a $j^{th}$ tier BS at $x_j$ if
\begin{equation}
\frac{P_{rk}   e^{-\frac{W\Delta_k}{P_{rk}}}{(\frac{P_{rk}}{\Delta_k})}^{\Psi_k}}
     {W^{\Psi_k-1}\Gamma(1-\Psi_k,\frac{W\Delta_k}{P_{rk}})}
>
\frac{P_{rj} e^{-\frac{W\Delta_j}{P_{rj}}}{(\frac{P_{rj}}{\Delta_j})}^{\Psi_j}}
     {W^{\Psi_j-1}\Gamma(1-\Psi_j,\frac{W\Delta_j}{P_{rj}})}\label{eq:lemma1_eq2}
\end{equation}
with the assumption that the interference from all other BSs and noise is fixed and denoted by residue $W$. Here $P_{ri}=P_{ri}(x_i)$
is the average received power from $i^{th}$ tier BS situated at $x_i$ for $i=j,k$ and $\Gamma(\cdot,\cdot)$ is the incomplete Gamma function. With the further assumption $W=0$, the above selection criterion can be simplified to
\begin{eqnarray}
P_{rk}\sqrt{\frac{(\Psi_k-1)}{\Delta_k }} &> & P_{rj}\sqrt{\frac{ (\Psi_j-1)}{\Delta_j}}.\label{eq:sl}
\end{eqnarray}
\end{lemma}

\begin{IEEEproof}
Using a similar idea as Lemma \ref{lemma:meansinr}, the mean SINR for BS at $x_k$ conditional on BS locations for the assumed case is given as
\begin{eqnarray*}
\mathbb{E}[\gamma_k | \phi]&=&P_{rk} \int_0^{\infty}{e^{-WT}\frac{1}{{(1+ T{P_j}{{\|x_j\|}^{-\alpha_j}})}^{\Psi_j}}}{\rm d}T.
\end{eqnarray*}
Substituting $W(1+{P_j}{{\|x_j\|}^{-\alpha_j}}T)={P_j}{{\|x_j\|}^{-\alpha_j}}y=P_{rj}y/\Delta_j$, the above expression simplifies to
\begin{eqnarray*}
\mathbb{E}[\gamma_k | \phi]&=& P_{rk} e^{W\Delta_j/P_{rj}}\frac{W^{\Psi_j-1}}{{(P_{rj}/\Delta_j)}^{\Psi_j}}\int\limits_{W\Delta_j/P_{rj}}^{\infty}{e^{-y}\frac{1}{{y}^{\Psi_j}}}{\rm d}y\\
&=&P_{rk} e^{W\Delta_j/a_j}\frac{W^{\Psi_j-1}}{{(P_{rj}/\Delta_j)}^{\Psi_j}}\Gamma\left(1-\Psi_j,\frac{W\Delta_j}{P_{rj}}\right),
\end{eqnarray*}
which leads to the first rule in the lemma. Further taking the  limit $W\rightarrow 0$, and using the fact that $\lim_{x\rightarrow 0}{\frac{\Gamma(-a,x)}{x^a}}=\frac{1}{a};a> 0$, we get
\begin{eqnarray*}
\mathbb{E}[\gamma_k]&=& P_{rk} e^{\frac{W\Delta_j}{P_{rj}}}\lim_{W\rightarrow 0}\frac{W^{\Psi_j-1}}{{(P_{rj}/\Delta_j)}^{\Psi_j}}\Gamma\left(1-\Psi_j,\frac{W\Delta_j}{P_{rj}}\right)\\
&=&\Delta_j \frac{P_{rk}}{P_{rj}}
\lim_{W\rightarrow 0}
{\left(\frac{W\Delta_j}{P_{rj}}\right)}^{\Psi_j-1}\Gamma\left(1-\Psi_j,\frac{W\Delta_j}{P_{rj}}\right)\\
&=&\frac{\Delta_j}{\Psi_j-1} \frac{P_{rk}}{P_{rj}}
\end{eqnarray*}
which leads to the second rule.
\end{IEEEproof}

\begin{figure}
\centering
\subfigure[]{\includegraphics[width=.9\columnwidth,trim=70 45 60 31,clip=true] {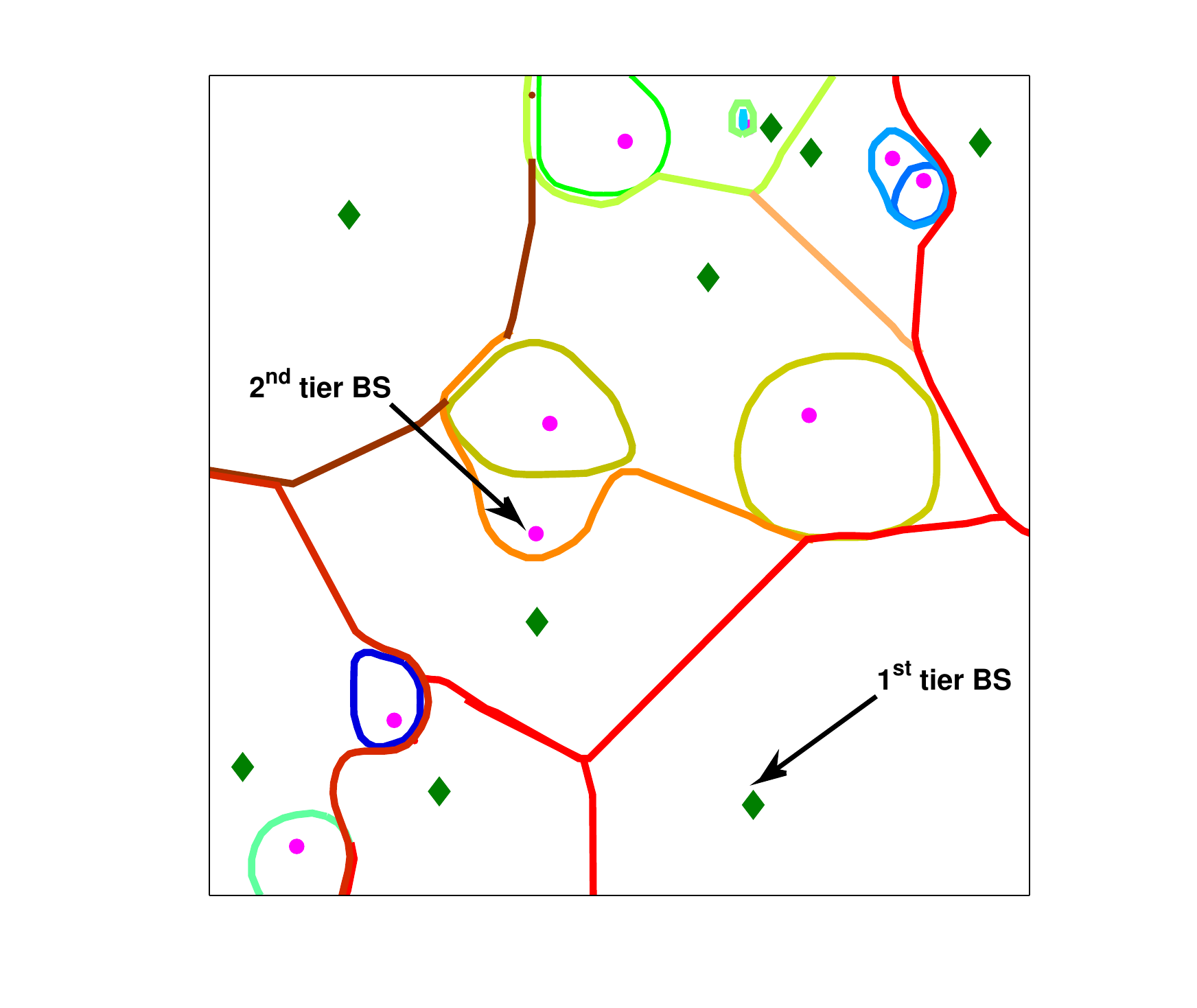}\label{fig:ECLP1}}
\subfigure[]{\includegraphics[width=.9\columnwidth,trim=70 45 60 31,clip=true]{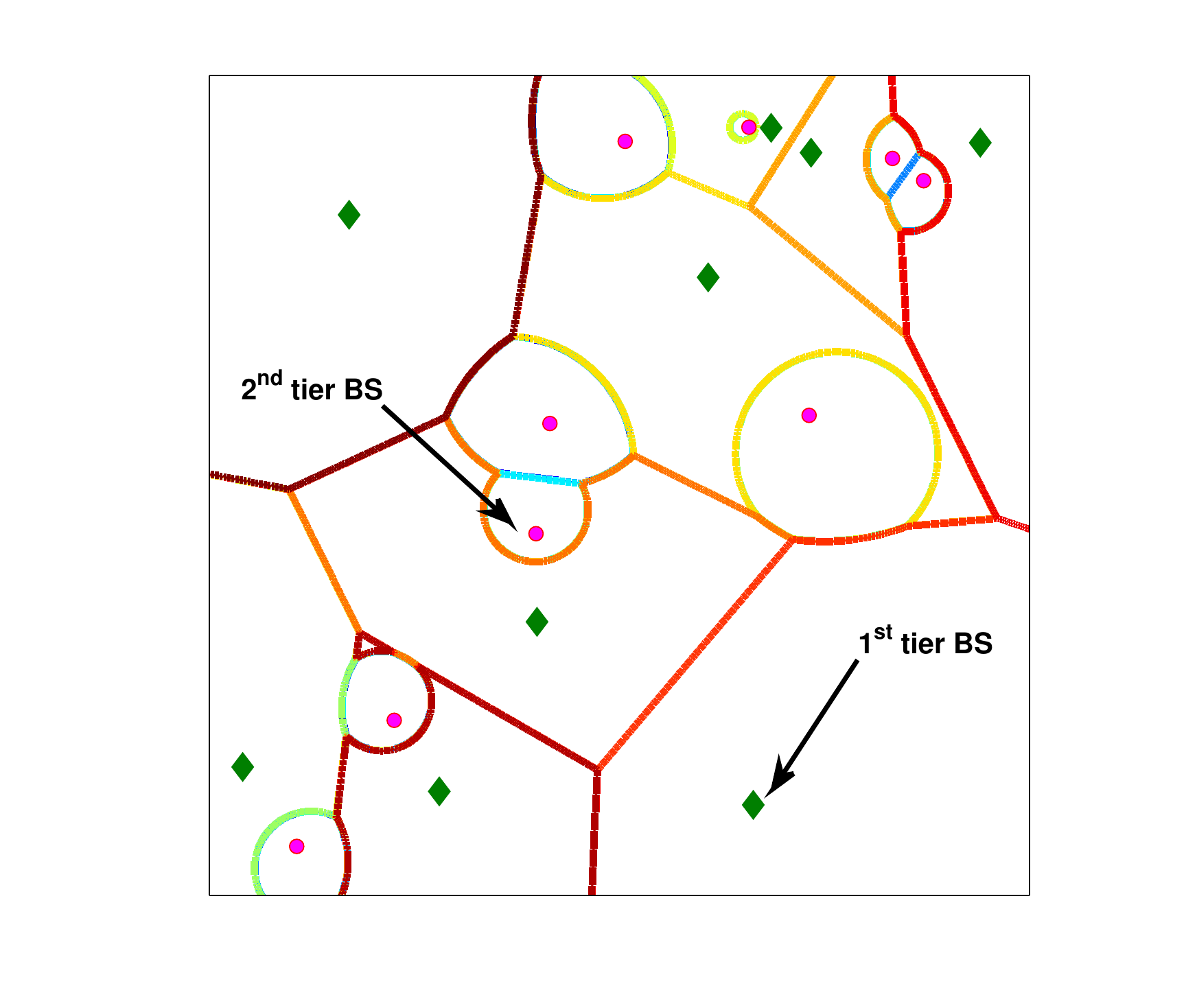}\label{fig:ECLP2}}
\caption{Cell Selection region for $P_1 = 5P_2, \lambda_1 = \lambda_2, \Psi= [3, 2], \Delta = [2, 1]$: (a) from simulation, where selection is based on highest mean SINR, and (b) from theory, where selection is based on Lemma \ref{lemma:meansinrw0}.
}
\label{fig:ECLregion}
\end{figure}

The above discussions provide enough evidence that a selection-bias based criterion is a meaningful option for cell selection in multi-antenna HetNets. Although in the previous lemma, we have taken $W$ to be a fixed constant which is generally not the case, still it indicates that taking the multiplicative form  of sufficient conditions, i.e., \eqref{Eq:SO3} is a better candidate function. As validated in the numerical results section, the bias value of $B_k = \sqrt{\frac{\Psi_k}{\Delta_k}}$ is surprisingly accurate for SINR, and hence coverage, maximization. Note, however, that this choice of selection bias is not necessarily optimal for other metrics considered in this paper, e.g., rate coverage. Therefore, to maintain generality, we derive all the results in terms of an arbitrary cell-selection bias term $B_k$, i.e., a user selects a nearest BS of $k^{th}$ tier if
\begin{align}
k = \arg \max_{j \in \mathcal{K}} B_j P_j\Delta_j {\|x_j\|}^{-\alpha_j}.\label{eq:modifybias}
\end{align}

\subsection{Association Region}
Under the selection rule discussed in the previous section, a typical user selects a BS that provides the maximum ``biased'' received power, where the bias values are tuned according to the metric that is being maximized. This creates exclusion regions around a typical user in which the interfering BSs can not lie. These exclusion regions are characterized in the following Lemma.

\begin{lemma}\label{lemma:associationregion}
A typical user selects $k^{th}$ tier BS located at a distance $d_k$ if the closest BSs of all the other tiers, located at distance $d_j, j\ne k$,   satisfy the following condition
\begin{eqnarray}
d_j &\ge&{\left({\widehat{P}_j}{\widehat{B}_j}{\widehat{\Delta}_j}\right)}^{\frac{1}{\alpha_j}}{d_k}^{\frac{1}{\widehat{\alpha}_j}},\label{eqlemma1}
\end{eqnarray}
 where $\widehat{P}_j=P_j/P_k$ and similarly for $\widehat{B}_j,\widehat{\Delta}_j$ and $\widehat{\alpha}_j$.
\end{lemma}
\begin{IEEEproof}
See Appendix \ref{Proof:associationregion}.
\end{IEEEproof}

Due to the nature of the cell selection rule, all the interfering BSs will satisfy \eqref{eqlemma1}. Lemma \ref{lemma:associationregion} shows that association regions are weighted Voronoi regions where weights are not just equal to the received power but also include a bias term which accounts for minimizing the interference. To justify our model, we compare it with a simulated mean SINR based selection model. First, the locations of the BSs are sampled from a PPP over a small spatial window. Then, the received SINR from each BS is computed for every point (on a grid) in the space according to appropriate fading distribution.
Fig. \ref{fig:ECLP1} shows the simulated coverage area where the user connects to the BS having best mean SINR (averaged over fading) conditioned on the given realization of BSs' positions whereas Fig. \ref{fig:ECLP2} shows the coverage area based on approximate modified bias association region for the same realization of BSs' locations. For this particular realization, the error region where the above two regions do not match is $0.3$ percentage of the total simulated area which is surprisingly accurate given the simple biasing-based approach used for the second plot.

\begin{figure}
\centering
\includegraphics[width=.9\columnwidth,trim=65 55 45 35,clip=true] {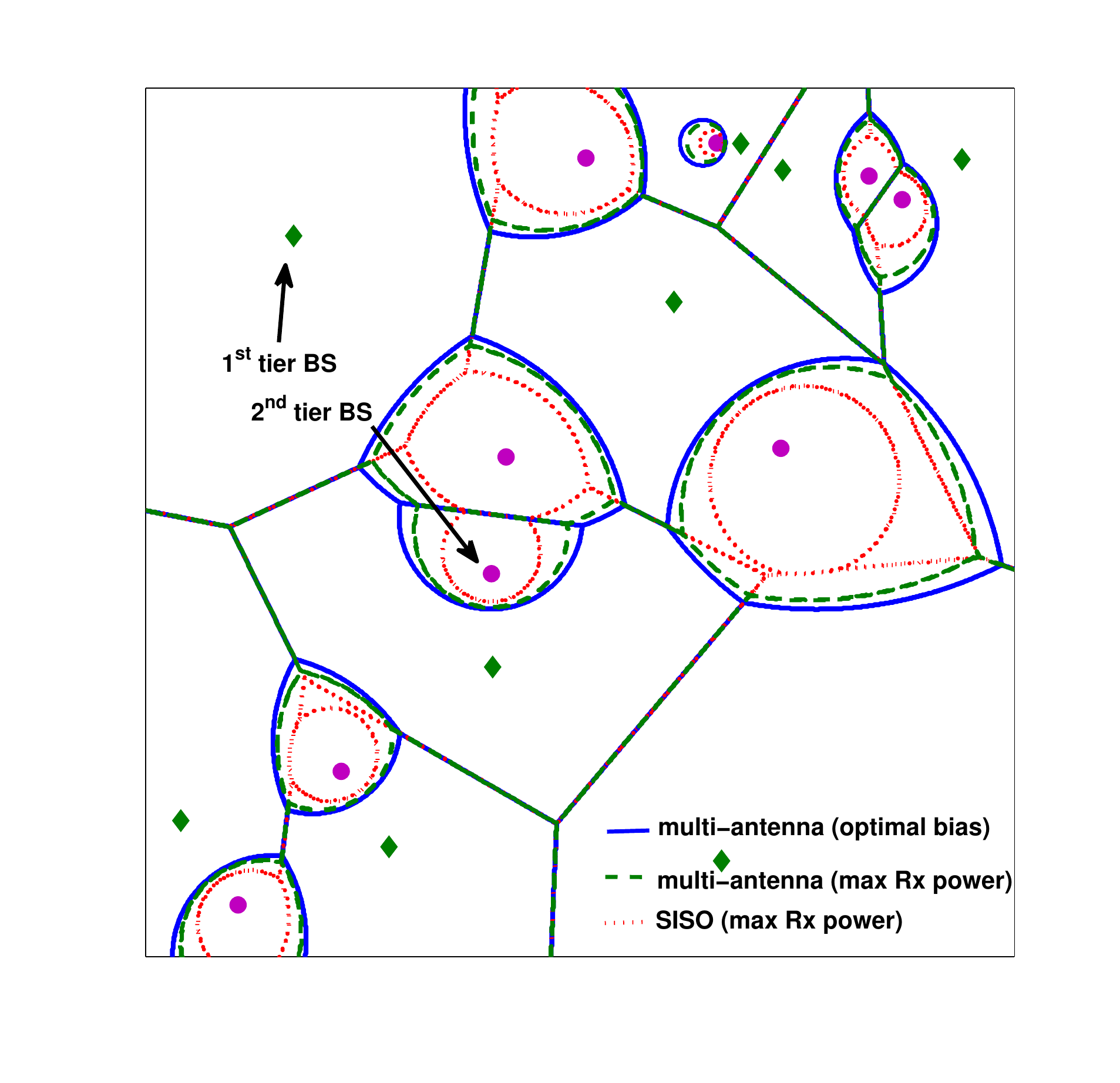}
\caption{The expanding of the coverage regions due to inherent biasing in two-tier multi-antenna HetNet with 	$\Psi_1 = 1; \Psi_2 = 6; P = [30; 1]; \Delta = [1; 3]$. In this hypothetical case, small cells (denoted by filled circle) use 8 antennas and schedule 3 users per resource block per BS. Macrocell BS (denoted by diamonds) have single antenna per BS.}
\label{fig:type2exapanding}
\end{figure}

Fig. \ref{fig:type2exapanding} compares the coverage regions of a SISO and multi-antenna HetNet with the same BS locations. For the multi-antenna HetNet, we consider two different cell selection rules: i) based on max received power, ii) based on max mean SINR. For the SISO HetNet, both these selection rules are exactly the same. As it can be seen that using multiple antennas and SDMA at the small cells includes a natural bias for these cells and therefore results in expansion of their coverage regions. This expansion naturally balances load across tiers and hence reduces the need for artificial bias to offload sufficient traffic to small cells compared to SISO HetNets. In the following example, we provide a useful insight into the selection of multi-antenna techniques for maximum expansion of the coverage regions of small cells with no external bias.

\begin{example} [Coverage expansion]
Consider a two-tier SISO HetNet consisting of macrocells and femtocells. Owing to their smaller transmit powers, femtocells have smaller coverage regions, which results in imbalanced load and degrades performance, especially in terms of rate~\cite{SinDhiJ2013}. However, the coverage regions of the femtocells can be naturally expanded by using multi-antenna transmission. For the sake of argument, assume that the number of antennas per femtocell in the new setup is $M=8$, while the macrocells still have $1$ antenna per BS. If cell-selection is based on maximum received power, optimal strategy is to maximize $\Delta$, which is achieved by single-user beamforming, i.e., $\Delta=8$ and $\Psi=1$. On the other hand, if the goal is to maximize the average SINR, the optimal strategy is the one that maximizes $\sqrt{\Delta \Psi}$. This is achieved under SDMA with $\Delta=5$ and $\Psi=4$. The sub-optimality of single-user beamforming for SINR maximization is counter-intuitive since single-user beamforming is mostly associated with range expansion of wireless links.
\end{example}

With these insights, we now derive the cell selection or cell association probability, which is the probability with which a typical user selects a $k^{th}$ tier BS.
\begin{lemma} \label{Lem:associationProb}
The probability that a typical user is associated with a $k^{th}$ tier BS is given as
\begin{eqnarray*}
A_k &=&2 \pi \lambda_k\int_{0}^{\infty}{
e^{-\pi \sum_{j}{ \lambda_j {\left(\widehat{P}_j\widehat{\Delta_j}\widehat{B}_j\right)}^{\frac{2}{\alpha}_j}{r}^{\frac{2}{\widehat{\alpha}_j}}}}r{\rm d}r}.
\end{eqnarray*}

If $\alpha_j=\alpha\ \forall j$, the above can be simplified to \begin{eqnarray*}
A_k &=&  \frac{ \lambda_k}{ \sum_{j}{ \lambda_j {\left(\widehat{P}_j\widehat{\Delta}_j\widehat{B}_j\right)}^{\frac{2}{\alpha}}}}.
\end{eqnarray*}
\end{lemma}
\begin{IEEEproof} The proof is the same as \cite[Lemma 1]{JoSanJ2012}, hence skipped. 
\end{IEEEproof}

\section{Coverage Probability}
Coverage probability is the probability that SINR at a typical user from the associated tier is above some threshold $T$ and is mathematically defined as
\begin{equation}\label{eq:Pcdefinition}
P_c=\sum_{k=1}^K\mathbb{P}[\gamma_k>T, k=\text{associated tier}].
\end{equation}
For the mean SINR association model, the coverage probability is given as
\begin{align}
{P}_c&=\sum_{k=1}^{K}\mathbb{E}_{\Phi_k}\left[\mathbbm{1}_{(\gamma(x_k)>T)}\mathbbm{1}_{(S_{x_k,\phi}\ge S_{y,\phi} \ \forall y\in \phi\backslash x_k)}\right]
\label{eq:Pc_theorem2}
\end{align}
where $\gamma_k$ and $\mathbb{E}[\gamma_k|\phi]$ are given by \eqref{expSINR} and \eqref{eq:meansinrdef} and $S_{y,\phi}=\mathbb{E}[\gamma(y)|\phi]$.

The above expression is quite complicated and will need the higher order factorial moments of a PPP. For the rest of this discussion, we will assume the selection bias-based model discussed in previous section. For this model, $S_k=P_{rk}B_k$ and we can further simplify the expression~\eqref{eq:Pc_theorem2} using iterative conditioning and get
\begin{align}
P_c&=&2\pi\sum_{k=1}^{K}\lambda_k\int_0^{\infty}\mathbb{P}\left(P_kh_{xk}\|x\|^{-\alpha_k}>T(N+I_{\Phi^{'}})\right)\cdot\nonumber\\
&&\ \  e^{-\sum_{j=1}^{K}\lambda_j\pi{(\widehat{P}_j\widehat{\Delta}_j\widehat{B}_j)}^{\frac{2}{\alpha_j}} x^{\frac{2}{\widehat{\alpha}_j}}}x{\rm d}x, \label{eq:outageprob}
\end{align}
where $\Phi^{'}$ is
\begin{equation}
\Phi^{'}=\cup_{k=1}^K{ \Phi_k \cap {B\left(0,{(\widehat{P}_j\widehat{\Delta}_j\widehat{B}_j)}^{\frac{1}{\alpha_j}} x^{\frac{1}{\widehat{\alpha}_j}}\right)}^c},
\end{equation}
where $B(0,r)$ is open ball around origin with radius $r$ and ${(.)}^c$ denotes the set complement. The Proof of \eqref{eq:outageprob} is given in Appendix \ref{Proofeqoutageprob}.


In the following subsections, we will compute the probability term inside the integral in \eqref{eq:outageprob}, which can be interpreted as the CCDF of SINR at a typical user associated with a $k^{th}$ tier BS located at $x_k\in \Phi_k$. Before going into further details, it is important to understand the resulting form of this term. Since SINR at a typical user depends only on the magnitude of $x_k$, i.e., the distance of BS $x_k \in \Phi_k$ from the origin, we will denote SINR by $\gamma_k(x_k)=\gamma_k(\|x_k\|)$ with slight abuse of notation. Letting $\|x_k\| = x$, the CCDF of SINR can be expressed as
\begin{eqnarray}
\mathbb{P}[\gamma_k(x) > T] &=& \mathbb{P}[P_k h_{xk}x^{-\alpha_k} > T (I+N)]\nonumber\\
&=&\mathbb{P}[h_{xk}>TP_k^{-1}x^{\alpha_k}(I+N)]\nonumber\\
&=&\sum_{i=0}^{\Delta_k-1}{\frac{1}{i!}\mathbb{E}\left[\left[-s(I+N)\right]^i e^{\left[-s(I+N)\right]}\right]}\nonumber\\
&=&\sum_{i=0}^{\Delta_k-1}{\frac{1}{i!}{(-s)}^i\frac{{\rm d}^i}{{\rm d}s^i}\left[\mathbb{E}\left[e^{-s(I+N)}\right]\right]}, \label{eqSINR1}
\end{eqnarray}
with $s=TP_k^{-1}x^{\alpha_k}$. As is clear from \eqref{eqSINR1}, we will not only need the Laplace transform of the interference but also its derivatives, which we compute in the following subsections.

\subsection{Laplace transform of noise plus Interference}
Since the typical user is associated with a $k^{th}$ tier BS located  at $x_k$ with  $\|x_k\|=x$, all the other BSs satisfy \eqref{eqlemma1} and are therefore located outside the ball of radius $r_j$, where
\begin{eqnarray*}
r_j &=&{\left(\widehat{P}_j\widehat{\Delta}_j\widehat{B}_j\right)}^{\frac{1}{\alpha_j}}{x}^{\frac{1}{\widehat{\alpha}_j}}.
\end{eqnarray*}
for all $j=1,2,\cdots K$.
\begin{theorem}
The Laplace transform of the noise plus interference from all BSs at a typical user associated with a $k^{th}$ tier BS located at a distance $x=\|x_k\|, x_k \in \Phi_k$, is given by \begin{align}
\mathcal{L}_{IN}(s)&=e^{-sN}\exp\left[-2\pi\sum_{j=1}^{K}\frac{\lambda_j}{\alpha_j}{(sP_j)}^{\frac{2}{\alpha_j}}\sum_{m=1}^{\Psi_j} \binom{\Psi_j}{m} \right.\nonumber
\\&
\left.
B'\left(\Psi_j+m-\frac{2}{\alpha_j},m+\frac{2}{\alpha_j},\frac{1}{1+\frac{sP_j{x}^{-\alpha_k}}{\left(\widehat{P}_j\widehat{\Delta}_j\widehat{B}_j\right)}}\right)\right],\nonumber\\
\end{align}
where $B'(a,b,c)$ is the complementary incomplete Beta function defined as
\begin{eqnarray*}
B'(a,b,z)&=&\int_{z}^{1}{u^{a-1}{(1-u)}^{b-1}{\rm d}u}.
\end{eqnarray*}

\end{theorem}
\begin{IEEEproof}
See Appendix \ref{AppendixProofInter} for the proof.
\end{IEEEproof}

\begin{corollary}
The Laplace transform of the noise plus interference at $s=TP_k^{-1}x^{\alpha_k}$ is given as
\begin{eqnarray*}
\mathcal{L}_{IN}(TP_k^{-1}x^{\alpha_k})
&=&e^{-TP_k^{-1}x^{\alpha_k}N}e^{\left[-\sum_{j=1}^{K}{\lambda_j{(T\widehat{P}_j)}^{\frac{2}{\alpha_j}}x^{\frac{2}{\widehat{\alpha}_j}}\calCfunc_j}\right],}
\end{eqnarray*}
where $\calCfunc_j$ is defined  as $\calCfunc_j\stackrel{\Delta}{=}$
\begin{align}
 &\frac{2\pi}{\alpha_j}\sum_{m=1}^{\Psi_j}{  \binom{\Psi_j}{m} 
B'\left(\Psi_j+m-\frac{2}{\alpha_j},m+\frac{2}{\alpha_j},\frac{1}{1+\frac{T}{\widehat{\Delta}_j\widehat{B}_j}}\right)}.\nonumber
 \end{align}
Further, if $\alpha_j=\alpha\ \forall j$, the Laplace transform of the noise plus interference at $s=TP_k^{-1}x^{\alpha}$ is 
\begin{eqnarray*}
\mathcal{L}_{IN}(TP_k^{-1}x^{\alpha}) &=&e^{-TP_k^{-1}Nx^{\alpha}-\left[\sum_{j=1}^{K}{\lambda_j{(T\widehat{P}_j)}^{\frac{2}{\alpha}}\calCfunc_j}\right]x^2}.
\end{eqnarray*}
\end{corollary}

\subsection{Derivatives of the Laplace Transform}
We now provide an expression for the derivative of the Laplace transform along with a brief sketch of the proof. The proof is based on the tools developed in \cite{DhiKouJ2013}.
\begin{theorem}\label{theorem:derivative}
The $n^{th}$ derivative  of the Laplace transform of noise plus interference (computed in section III-A) is given as
\begin{align}
&\frac{{\rm d}^n}{{\rm d}s^n}\mathcal{L}_{IN}(s)=\mathcal{L}_{IN}(s)\sum_{\bar{m}\in M}{C(\bar{m})}\nonumber\\
&{
\cdot\prod_{l=1}^{n}{{\left(-N\mathbbm{1}_{l=1}+2\pi\sum_{j=1}^{K}{{(-1)}^{l}D_j(l)P_j^l
{(sP_j)}^{\frac{2}{\alpha_j}-l}
}\right)}^{m_l}}},\nonumber
\end{align}
 where \begin{eqnarray*}
M(n)&=&\{\bar{m}=(m_1,m_2,\cdots m_n)^T:\sum_{i=1}^{n}{im_i}=n\}\\
C(\bar{m})&=&\frac{n!}{\prod_i{\left(m_i!{(i!)}^{m_i}\right)}}\\
D_j(l)&=&\frac{\lambda_j}{\alpha_j}\frac{(\Psi_j+l-1)!}{(\Psi_j-1)!}B'\left(\Psi_j+\frac{2}{\alpha_j},l-\frac{2}{\alpha_j},u_j\right)\\
u_j&=&\frac{1}{1+\frac{sP_j{x}^{-\alpha_k}}{\left(\widehat{P}_j\widehat{\Delta}_j\widehat{B}_j\right)}}.
\end{eqnarray*}
\end{theorem}
\begin{IEEEproof}
Proof is given in Appendix \ref{AppendixProofInterDeriv}.\end{IEEEproof}

For the  interference limited case, the above expression can be simplified further and is given in the following Corollary.
\begin{corollary}
If $\alpha_j=\alpha \ \forall j$ and noise $N=0$, for $s=TP_k^{-1}x^{\alpha}$, 
\begin{align}
\frac{{\rm d}^n\mathcal{L}_I(s)}{{\rm d}s^n}&=e^{-\left[\sum_{j=1}^{K}{\lambda_j{(T\widehat{P}_j)}^{\frac{2}{\alpha}}\calCfunc_j}\right]x^2} \cdot \nonumber\\
&\sum_{\bar{m}\in M}{C(\bar{m})x^{-n\alpha+2\sum{m_l}}{(-1)}^nF(\bar{m})},\ \ \
\end{align}
where 
\begin{eqnarray*}
   F(\bar{m}) &\stackrel{\Delta}{=}& \frac{{(2\pi)}^{\sum{m_l}}}
    {{(TP_k^{-1})}^{-\frac{2}{\alpha}\sum{m_l}+n}}
\prod_{l=1}^{n}{\left(
        \sum_{j=1}^{K}
                {
                D_j(l)
                P_j^{\frac{2}{\alpha}}
            }
    \right)
    }^{m_l}
\end{eqnarray*}
with $u_j=\frac{1}{1+\frac{T}{\widehat{\Delta}_j\widehat{B}_j}}$.
\end{corollary}
\begin{IEEEproof}
The result follows simply by substituting $\alpha_j=\alpha,\widehat{\alpha}_j=1,N=0$, taking terms containing $x$ out of the product and using the fact that $\prod_{l=1}^{n}{{(x^{t-l})}^{m_l}}=x^{t(\sum m_l)-n}$.
\end{IEEEproof}

\subsection{SINR distribution}
Using  \eqref{eqSINR1} and the results derived in the previous two subsections, we can now compute the CCDF of SINR, which is given by the following Lemma.
\begin{lemma} \label{thm:SINRdist}
The CCDF of SINR at a typical user associated with the $k^{th}$ tier BS located at a distance $x$ from the user is 
$\mathbb{P}[\gamma_k(x) > T]=$
\begin{align}
&\sum_{n=0}^{\Delta_k-1}\frac{{(-1)}^nT^nx^{n\alpha_k}}{n!P_k^n}
e^{-\frac{Tx^{\alpha_k}N}{P_k}}\cdot\nonumber\\
& e^{\left.-\sum_{j=1}^{K}{\lambda_j{(T\widehat{P}_j)}^{\frac{2}{\alpha_j}}x^{\frac{2}{\widehat{\alpha}_j}}
\calCfunc_j}\right.}\sum_{\bar{m}\in M}{C(\bar{m})}\cdot \nonumber\\
&\prod_{l=1}^{n}{{\left(N\mathbbm{1}_{l=1}+2\pi\sum_{j=1}^{K}{{(-1)}^l
D_j(l)
P_j^l
{(T\widehat{P}_j)}^{\frac{2}{\alpha_j}-l}\frac{x^{\frac{2}{\widehat{\alpha}_j}}}{x^{l\alpha_k}}
}\right)}^{m_l}}
\end{align}
with $u_j=\frac{1}{1+\frac{T}{\widehat{\Delta}_j\widehat{B}_j}}$.
\end{lemma}

\begin{IEEEproof}
Proof is given in Appendix \ref{AppendixProofSINRDist}.
\end{IEEEproof}

\begin{corollary}
If we assume $\alpha_j=\alpha \ \forall j$  and interference limited case ($N=0$), the CCDF of SINR at typical user is given by $P[\gamma_k(x) > T]=$
\begin{align*}
&\sum_{n=0}^{\Delta_k-1}\frac{1}{n!}{(-TP_k^{-1}x^{\alpha})}^n
e^{-\left[\sum_{j=1}^{K}{\lambda_j{(T\widehat{P}_j)}^{\frac{2}{\alpha}}\calCfunc_j}\right]x^2}\cdot\\
&\sum_{\bar{m}\in M}{C(\bar{m})x^{-n\alpha+2\sum{m_l}}{(-1)}^nF(\bar{m})}\\
&=\sum_{n=0}^{\Delta_k-1}\frac{1}{n!}{(TP_k^{-1})}^n\cdot\\
&\sum_{\bar{m}\in M}C(\bar{m})F(\bar{m}) e^{-\left[\sum_{j=1}^{K}{\lambda_j{(T\widehat{P}_j)}^{\frac{2}{\alpha}}\calCfunc_j}\right]x^2}x^{2\sum{m_l}}.
\end{align*}
\end{corollary}

\subsection{Coverage Probability}
Substituting the SINR CCDF given by Theorem~\ref{thm:SINRdist} in \eqref{eq:outageprob}, we can compute the coverage probability, which is given by the following Theorem.

\begin{theorem}
Coverage probability of a typical user is $P_c=$
\begin{align*}
&\sum_{k=1}^K{2\pi \lambda_k}\int_{0}^{\infty}
\sum_{n=0}^{\Delta_k-1}\frac{1}{n!}{(-TP_k^{-1}x^{\alpha_k})}^n e^{-TP_k^{-1}x^{\alpha_k}N}\cdot\\
& e^{-\sum_{j=1}^{K}{\lambda_j{(T\widehat{P}_j)}^{\frac{2}{\alpha_j}}x^{\frac{2}{\widehat{\alpha}_j}}\calCfunc_j}}
\sum_{\bar{m}\in M}{C(\bar{m})}\cdot\\
& \prod_{l=1}^{n}{{\left(N\mathbbm{1}_{l=1}+2\pi\sum_{j=1}^{K}{
{(-1)}^l
D_j(l)
P_j^l
{(T\widehat{P}_j)}^{\frac{2}{\alpha_j}-l}\frac{x^{\frac{2}{\widehat{\alpha}_j}}}{x^{l\alpha_k}}
}\right)}^{m_l}}\cdot\\
&xe^{
-\pi \sum_{j=1}^{K}{\lambda_j{(\widehat{P}_j\widehat{\Delta}_j\widehat{B}_j)}^{\frac{2}{\alpha_j}}x^{\frac{2}{\widehat{\alpha}_j}}}
}{\rm d}x
\end{align*} with $u_j=\frac{1}{1+\frac{T}{\widehat{\Delta}_j\widehat{B}_j}}$.
\end{theorem}

\begin{corollary}
If we assume $\alpha_j=\alpha \ \forall j$ and interference limited case ($N=0$), the coverage probability of a typical user is given as $P_c=$
\begin{align}
&\Scale[0.97]{\sum_{k=1}^K{\pi \lambda_k}{}
\sum_{n=0}^{\Delta_k-1}\frac{1}{n!}{(TP_k^{-1})}^n} \cdot\label{Eq:CoverageFinal}\\
&\Scale[0.97]{\sum_{\bar{m}\in M} \frac{C(\bar{m})F(\bar{m})\Gamma(\sum{m_l}+1)}
{\left[\sum_{j=1}^{K}{\lambda_j{(T\widehat{P}_j)}^{\frac{2}{\alpha}}\calCfunc_j}+\pi \sum_{j=1}^{K}{\lambda_j{(\widehat{P}_j\widehat{\Delta}_j\widehat{B}_j)}^{\frac{2}{\alpha}}}\right]^{\sum{m_l}+1}}}\nonumber
\end{align}
\end{corollary}

\begin{IEEEproof}
See Appendix \ref{AppendixProofCov}.
\end{IEEEproof}

From \eqref{Eq:CoverageFinal}, we can observe that probability of coverage is no longer scale invariant in a multi-antenna HetNet even for the interference limited case. Coverage probability for a user associated with the tier $k$ is given as
\begin{align*}
P_{ck}&=P[\gamma_k>T|k=\text{associated tier}]\\
&=\frac{P[\gamma_k>T,k=\text{associated tier}]}{A_k},
\end{align*}
which is a decreasing function of  BS intensities $\lambda_j$'s $j\ne k$ of other tiers. We also observe that $P_{ck}$ is a decreasing function of $\Psi_j$'s of all the tiers.

\section{Rate Coverage}
In this section, we focus on the downlink rate achievable by a typical user. In particular, we derive the CCDF of downlink rate, which can be equivalently defined as rate coverage, i.e., the probability that the downlink rate achievable at a typical user is greater than a predefined target. This section generalizes the main ideas developed in \cite{SinDhiJ2013} for single-antenna HetNets to multi-antenna HetNets. Following the same setup as \cite{SinDhiJ2013}, we assume that each $k^{th}$ tier BS has same time-frequency resources $W_k$, which are equally distributed among all the users served by a given BS. Further assume that the $k^{th}$ tier BS that serves the typical user located at the origin, termed {\em tagged BS}, allocates $\mathcal{O}_k \leq W_k$ to each user, including the typical user. Therefore, the instantaneous rate $R_k$ achievable by a typical user when it connects to a $k^{th}$ tier BS is
\begin{align}
R_k &= \mathcal{O}_k\log_2{(1+\gamma_k(x_k))}.
\label{eq:R_k_main_def}
\end{align}

As discussed in detail in \cite{SinDhiJ2013}, the effective time-frequency resources $\mathcal{O}_k$ allocated to a typical user depend upon the number of users, equivalently load, served by the tagged BS, which is a random variable due to the random locations and hence the coverage areas of each BS. However, as argued in \cite{SinDhiJ2013} and verified further in~\cite{DhiAndJ2014}, approximating this load for each tier with its respective mean does not compromise the accuracy of results. Using results from \cite{SinDhiJ2013}, the mean load served by the tagged BS from $k^{th}$ tier can be approximated as $N_k=1+\frac{1.28\lambda_uA_k}{\lambda_k}$, where $\lambda_u$ is the density of the users and $A_k$ is the association probability given by Lemma~\ref{Lem:associationProb}. Note that the effect of multi-antenna transmission is captured in $A_k$. Now since each $k^{th}$ tier BS can schedule $\Psi_k$ users in a single resource block, total available (time-frequency) resource allocated to each user is
\begin{align}
\mathcal{O}_k&=\frac{W_k}{N_k/\Psi_k}.
\label{eq:O_k}
\end{align}

Combining the SINR distribution derived in the previous section and the average load result discussed above, the rate CCDF, equivalently rate coverage, can be derived on the same lines as \cite{SinDhiJ2013}. For easier exposition, we first derive the rate CCDF conditional on the serving BS being in the $k^{th}$ tier. The result is given by the following Theorem.
\begin{theorem}
The rate coverage for random selected user associated with $k^{th}$ tier is given by $\mathcal{R}_k=$
\begin{align*}
&\mathbb{P}[R_k>\rho]=\frac{2\pi \lambda_k}{A_k}\int_{0}^{\infty}
\sum_{n=0}^{\Delta_k-1}\frac{1}{n!}{(tP_k^{-1}x^{\alpha_k})}^n\\
&e^{-tP_k^{-1}x^{\alpha_k}N-\sum_{j=1}^{K}{\lambda_j{(t\widehat{P}_j)}^{\frac{2}{\alpha_j}}x^{\frac{2}{\widehat{\alpha}_j}}\calCfunc_j}}
\sum_{\bar{m}\in M}{C(\bar{m})}\cdot\\
&\prod_{l=1}^{n}{{\left(N\mathbbm{1}_{l=1}+2\pi\sum_{j=1}^{K}{
{(-1)}^l
D_j(l)
P_j^l
{(t\widehat{P}_j)}^{\frac{2}{\alpha_j}-l}\frac{x^{\frac{2}{\widehat{\alpha}_j}}}{x^{l\alpha_k}}
}\right)}^{m_l}}\cdot\\
&  xe^{
-\pi \sum_{j=1}^{K}{\lambda_j{(\widehat{P}_j\widehat{\Delta}_j\widehat{B}_j)}^{\frac{2}{\alpha_j}}x^{\frac{2}{\widehat{\alpha}_j}}}
}{\rm d}x,
\end{align*}
with $u_j=\frac{1}{1+\frac{t}{\widehat{\Delta}_j\widehat{B}_j}}$ and $t=2^{\frac{\rho N_k}{(W_k\Psi_k)}}-1$ and $N_k=1+\frac{1.28\lambda_uA_k}{\lambda_k}$ is the mean number of users served by the tagged BS.
\end{theorem}
\begin{IEEEproof}
The proof is similar to \cite{SinDhiJ2013}, hence omitted. Note that the only difference is the presence of $\Psi_k$, which depends upon the multi-antenna transmission. Putting $\Psi_k=1$ specializes this result to SISO HetNets, discussed in detail in \cite{SinDhiJ2013}.
\end{IEEEproof}
It is worth highlighting that although the conditional rate coverage computed above uses the mean load approximation for the tagged BS, we can easily incorporate the distribution of user load (see \cite{SinDhiJ2013}), which is skipped to avoid repetition. The rate coverage $R_c=\mathbb{P}[R>\rho]$ can now be computed as weighted sum of rate coverage of $i^{th}$ tier weighted by association probability of the respective tier
\begin{align}
R_c=\mathbb{P}[R>\rho]&=\sum_{i=1}^K{A_k \mathbb{P}[R_k>\rho]}.
\end{align}

Before concluding this section, it is important to note that a higher value of $\Psi_k$ means more users share the same time-frequency resources, which in turn means that each user gets higher chunk of $W_k$. This is also evident from~\eqref{eq:O_k}, where $\mathcal{O}_k \propto \Psi_k$. But since  coverage probability is a decreasing function of $\Psi_j$'s , the overall rate coverage expression represents a trade-off between available resources and SINR. 

\section{Numerical Results}

In this section, we validate our analysis and provide key design insights for multi-antenna HetNets. Before discussing the results, we briefly describe the simulation procedure. We choose a large spatial window and generate $K$ independent PPPs with the given densities. For every realization, a typical user is assumed at the origin and we select a serving BS according to the selection criteria \eqref{eq:modifybias}. Let this BS belongs to tier $i$. After this, fading random variable $h_{xi}$ is generated for the selected BS according to $\Gamma(\Delta_i,1)$ distribution and fading random variables $g_{xj}$ are generated for remaining BSs according to $\Gamma(\Psi_j,1)$.
A user is said to be in coverage if the SINR (or SIR in no-noise case) from the selected BS is greater than the target. The coverage probability is finally computed by averaging the indicator of coverage over sufficient realizations of the point process.

For concreteness, we restrict our simulation results to a two-tier HetNet with no noise, with the first tier denoting macrocells and the second denoting small cells, e.g., femtocells. With slight overloading of notation, the parameters are denoted by arrays, {e.g.}, $\lambda=[150, 300]$ means $\lambda_1=150,\lambda_2=300$. Following three antenna configurations are considered:

\noindent {\em 4-2 antenna configuration}: Macrocells have 4 antennas per BS, while femocells have 2 antennas per BS.

\noindent {\em 2-1 antenna configuration}: Macrocells have 2 antennas per BS, while femocells have 1 antenna per BS.

\noindent {\em SISO configuration}: All the BSs have single antenna.

In terms of multi-antenna transmission techniques, we restrict our attention to the following two techniques:

\noindent {\em Single-user beamforming (SUBF)}: where each BS serves single user per resource block, i.e.,  $\Delta_i=M_i,\Psi_i=1$.

\noindent {\em Full spatial division multiplexing (SDMA)}: where each $i^{th}$ tier BS serves $M_i$ users per resource block, i.e., $\Delta_i=1,\Psi_i=M_i$.

Note that when we combine these two multi-antenna transmission schemes with the three different antenna configurations discussed above, we already have {\em 7 simulation cases} to consider. We list these cases below for ease of exposition:

\noindent {\em 4-2 antenna configuration}
    \begin{itemize}
    \item {\em Case 1.} Both tiers use SUBF. 
    \item {\em Case 2.} Both tiers use SDMA.
    \item {\em Case 3.} First tier uses SDMA and other SUBF. 
    \item {\em Case 4.} First tier uses SUBF and other SDMA. 
    \end{itemize}
\noindent {\em 2-1 antenna configuration}
    \begin{itemize}
    \item {\em Case 5.} First tier uses SUBF. 
    \item {\em Case 6.} First tier uses SDMA.  
\end{itemize}
\noindent {\em SISO configuration} 
    \begin{itemize}
	\item {\em Case 7.} Both tiers use SISO.  
    \end{itemize}

\begin{figure}
\centering
\includegraphics[width=\columnwidth,trim=4 0 8 24,clip=true]{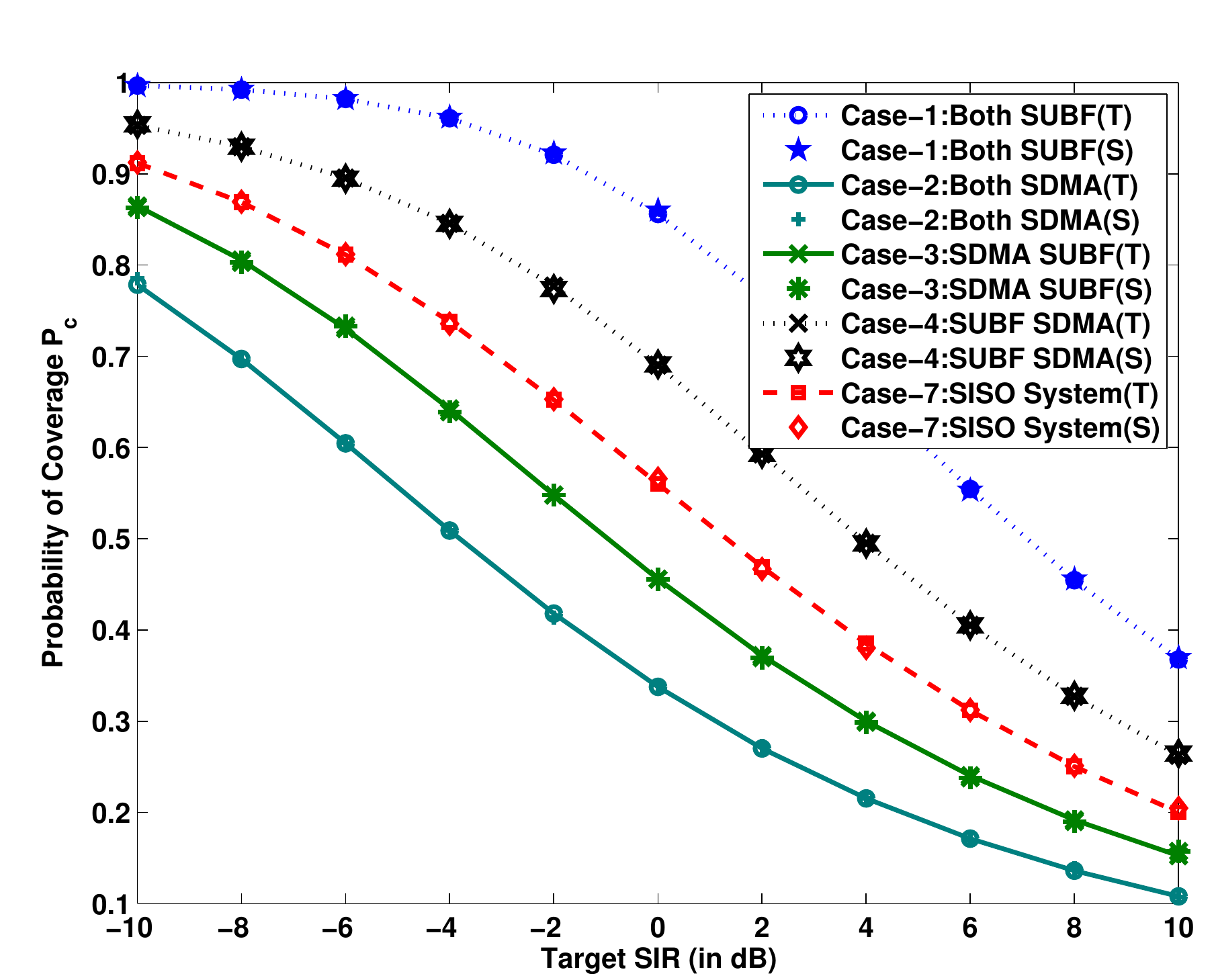}
\caption{Coverage probability of a two-tier HetNet with $\alpha=4, \lambda=[150,300],P_1=5P_2$ for 4-2 antenna configuration with SUBF, SDMA techniques and SISO system. T and S respectively denote theoretical and simulation results. Selection bias is $\sqrt{\Psi_j\Delta_j}$.}
\label{fig:PcAll}
\end{figure}

\begin{figure}
\centering
\includegraphics[width=\columnwidth,trim=4 0 8 24,clip=true]{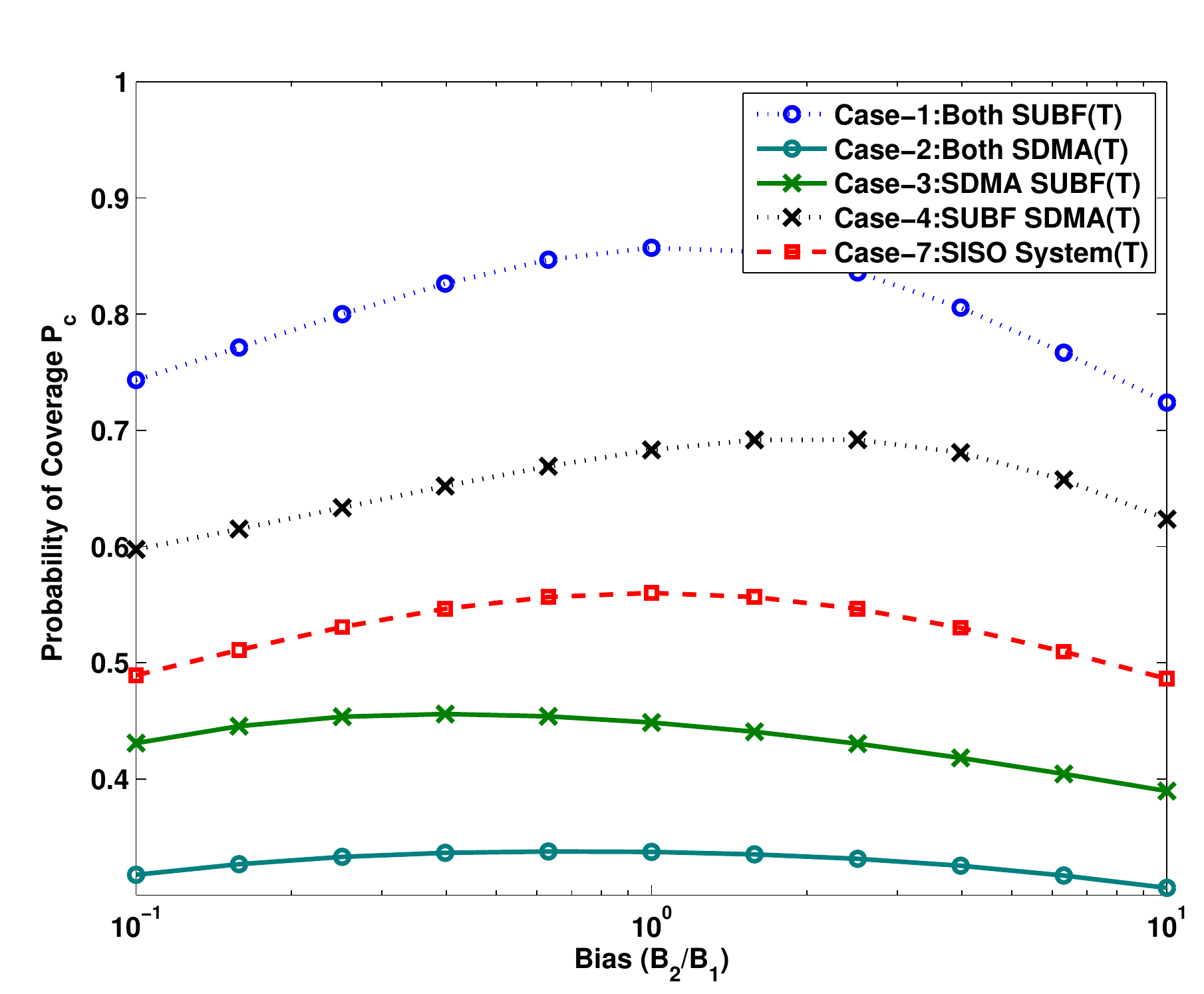}
\caption{Coverage probability versus relative bias $B_2/B_1$ in a two-tier HetNet with $\alpha=4, \lambda=[150,300],P_1=5P_2$ for SIR target $0$ dB  for {\em 4-2 antenna configuration}.}
\label{fig:BiasOptThreeruns}
\end{figure}

\begin{figure}
\centering
\includegraphics[width=\columnwidth]{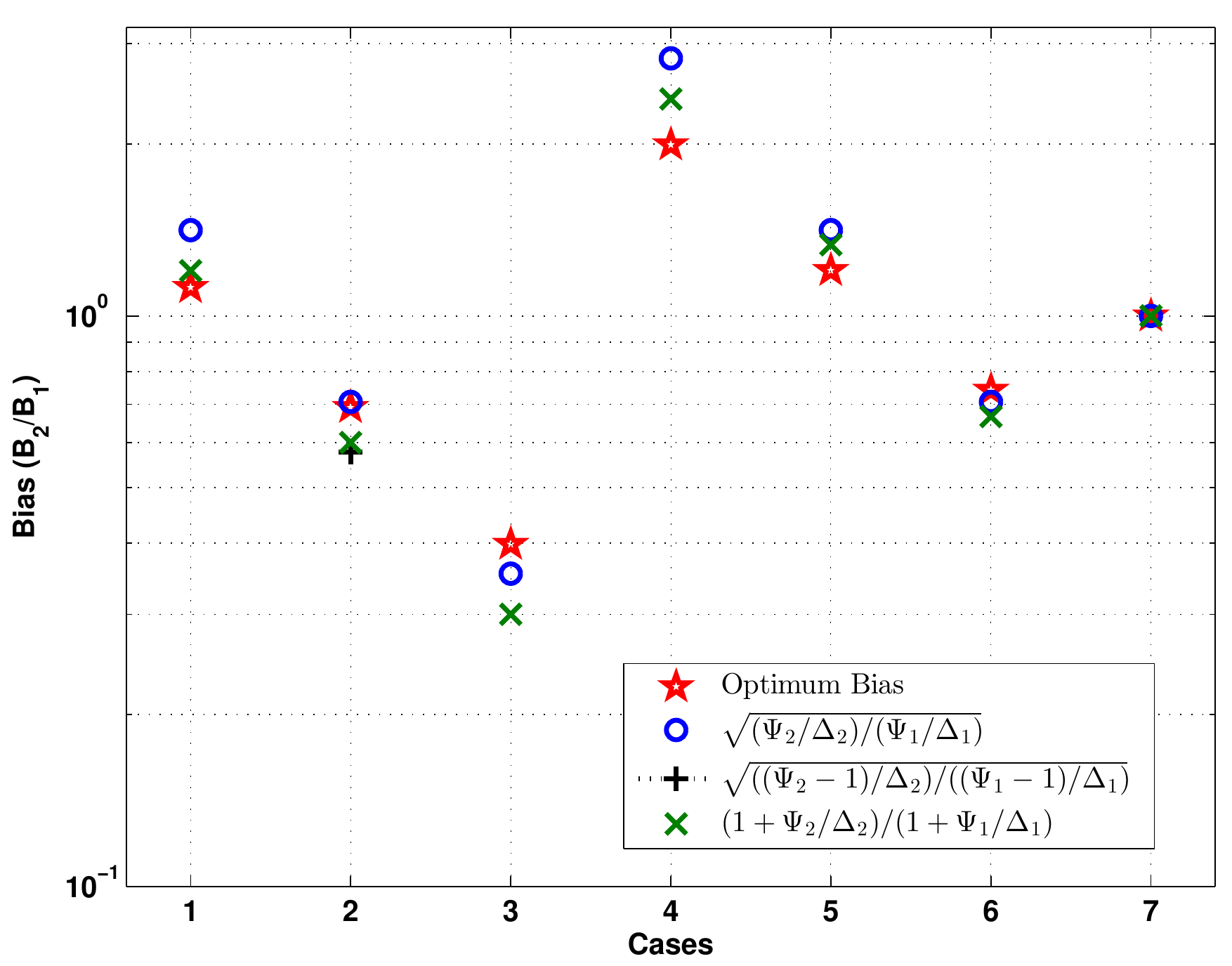}
\caption{Comparison of different candidates functions with numerically optimized bias for  all 7 simulation cases ($\alpha= 4$, target SIR = $0$ dB, $\lambda=[150,300],P_1=5P_2$).}
\label{fig:cand}
\end{figure}

\subsection{Coverage Probability}
Fig. \ref{fig:PcAll} shows the probability of coverage for cases 1-4 corresponding to {\em 4-2 antenna configuration} and case 7 corresponding to the {\em SISO configuration}. It can be seen that case 1, where both the tiers perform SUBF, results in the highest coverage. This result is consistent with~\cite{DhiKouJ2013}, where SUBF was shown to provide highest coverage under a slightly different cell selection model. On the other hand, SDMA performs worse than SISO because the effective fading gain from interfering BSs increases in mean  and thus causes stronger interference whereas the effective fading gain of the serving link remains the same as the SISO case. The other intermediate cases, where one tier performs SUBF and the other performs SDMA, fall in between these two extremes. 
\begin{figure}
\centering
\includegraphics[width=\columnwidth]{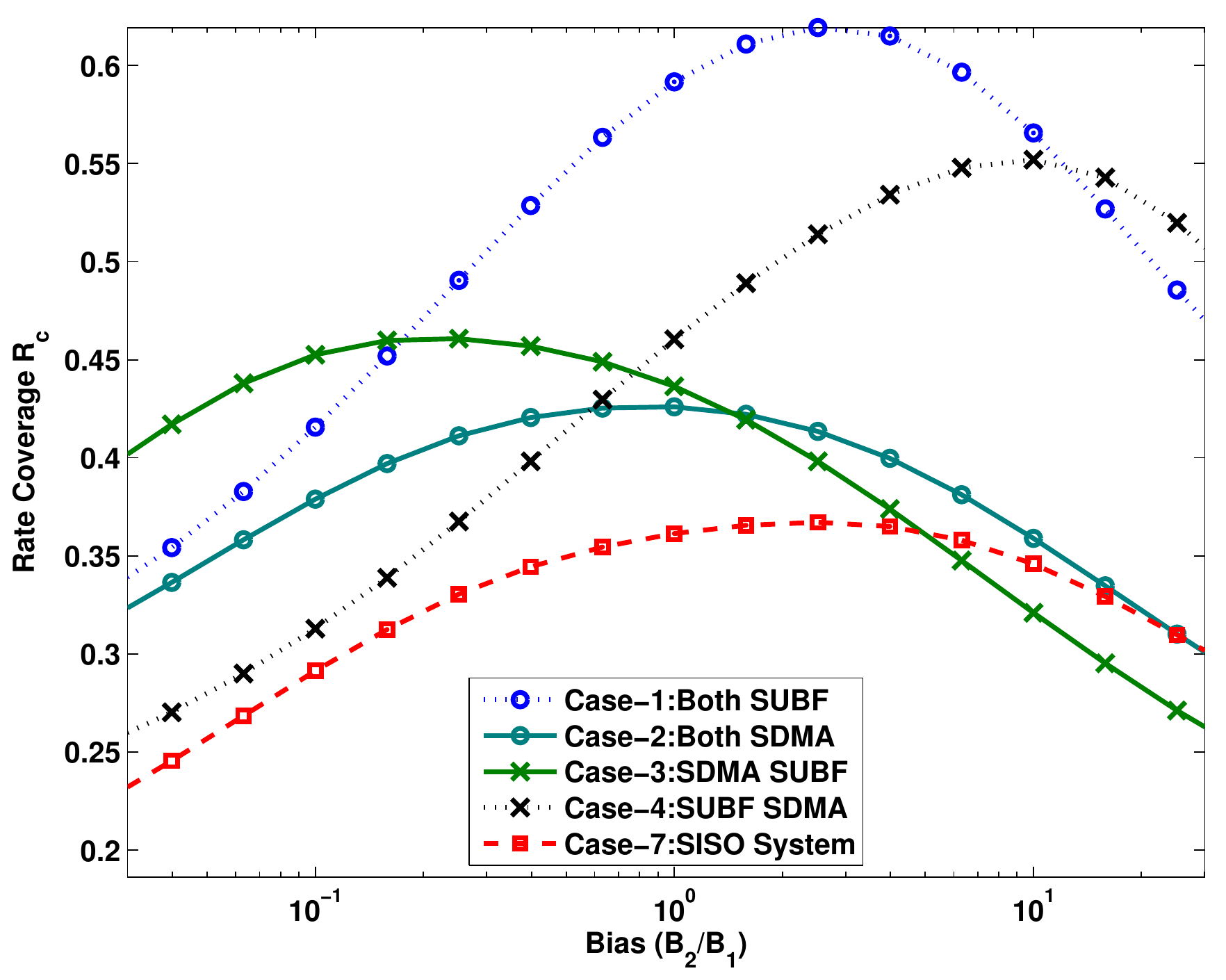}
\caption{Rate coverage probability versus relative bias $B_2/B_1$ in a two-tier HetNet with $\alpha=4, \lambda=[150,300],P_1=5P_2=50$ for rate threshold $1$ bps/Hz.}
\label{fig:BiasOptThreeruns_RC}
\end{figure}

\begin{figure}
\centering
\includegraphics[width=\columnwidth]{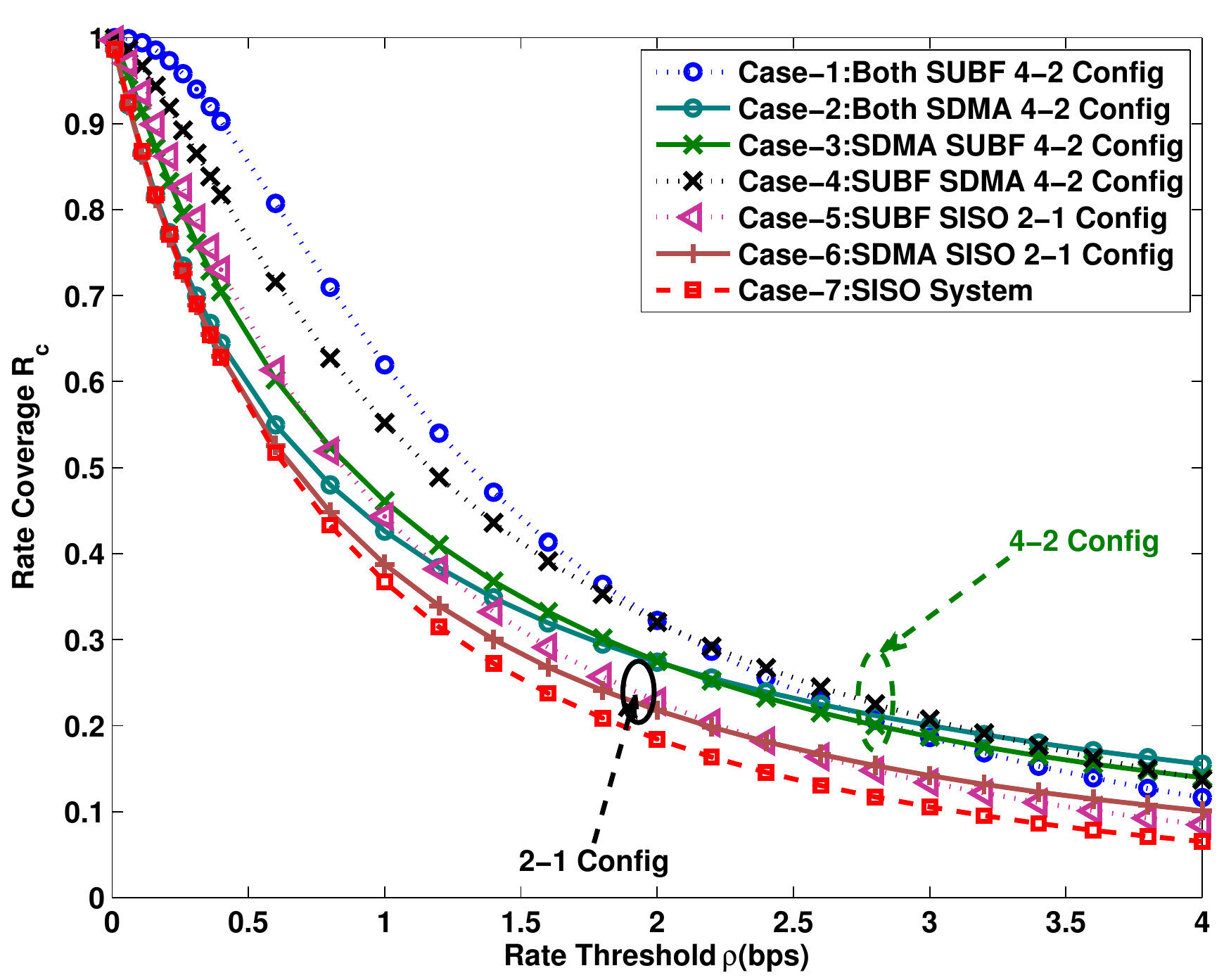}
\caption{Rate coverage with optimum bias $b=B_2/B_1$ in a two-tier HetNet with $\alpha=4,  \lambda=[150,300],P_1=5P_2$ for all 7 simulation cases. }
\label{fig:BiasOptimized_ratecoverage}
\end{figure}

\begin{figure}
\centering
\includegraphics[width=\columnwidth]{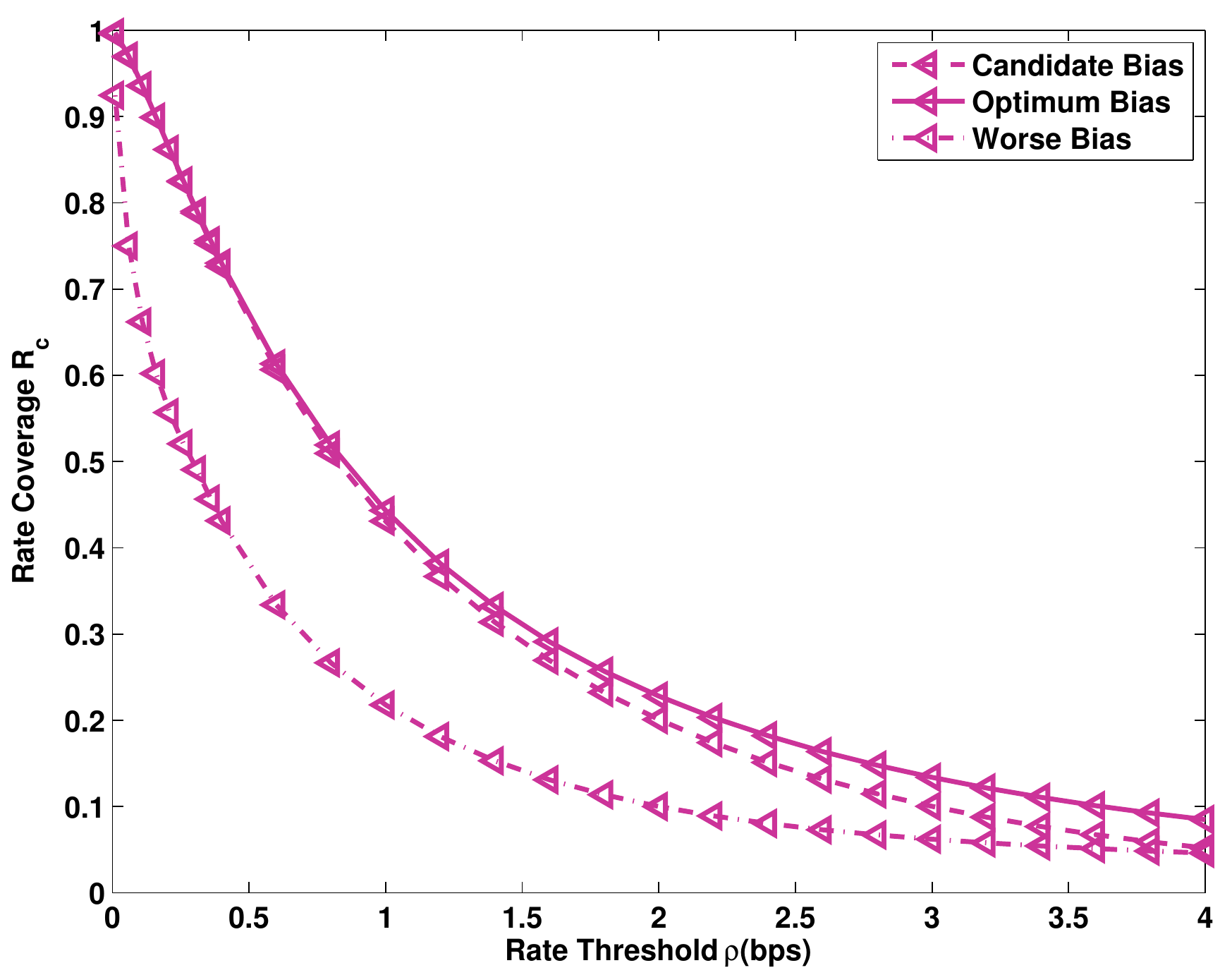}
\caption{Comparison of rate coverage with optimum bias $b=B_2/B_1$, candidate bias $B_i=\sqrt{\Psi_i\Delta_i}$ and worst bias in the range $[0.01,100]$ in a two-tier HetNet with $\alpha=4,  \lambda=[150,300],P_1=5P_2=50$ for simulation case 5.}
\label{fig:BiasOptimized_ratecoverage_comparison}
\end{figure}

\begin{figure}
\centering
\includegraphics[width=\columnwidth]{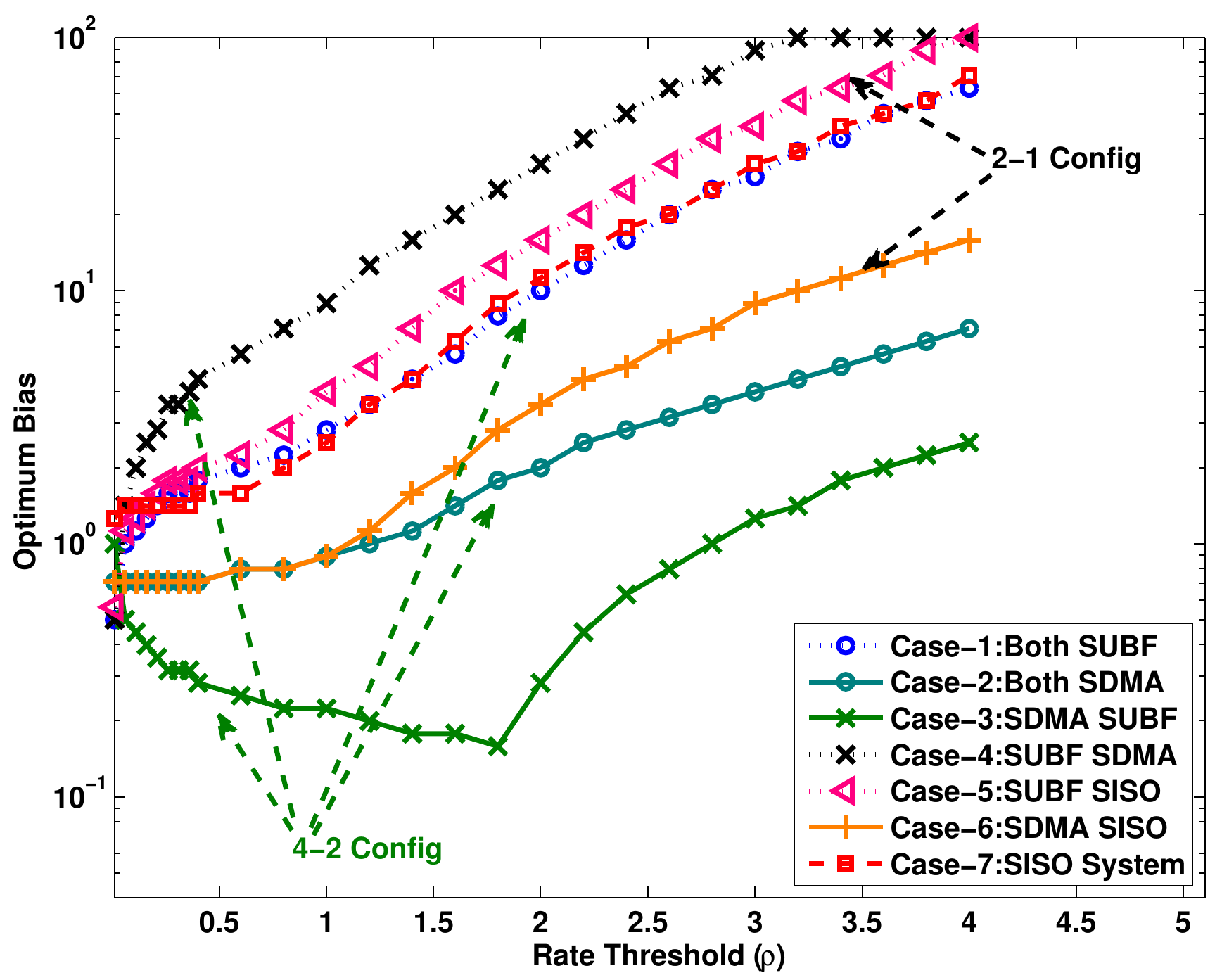}
\caption{Optimum bias $b=B_2/B_1$ for maximizing rate coverage probability versus target rate threshold  in a two-tier HetNet with $\alpha=4,  \lambda=[150,300],P_1=5P_2=50$ for different simulation cases.}
\label{fig:BiasOptimized_ratecoverage_biasopt}
\end{figure}

\begin{figure}
\centering
\includegraphics[width=\columnwidth]{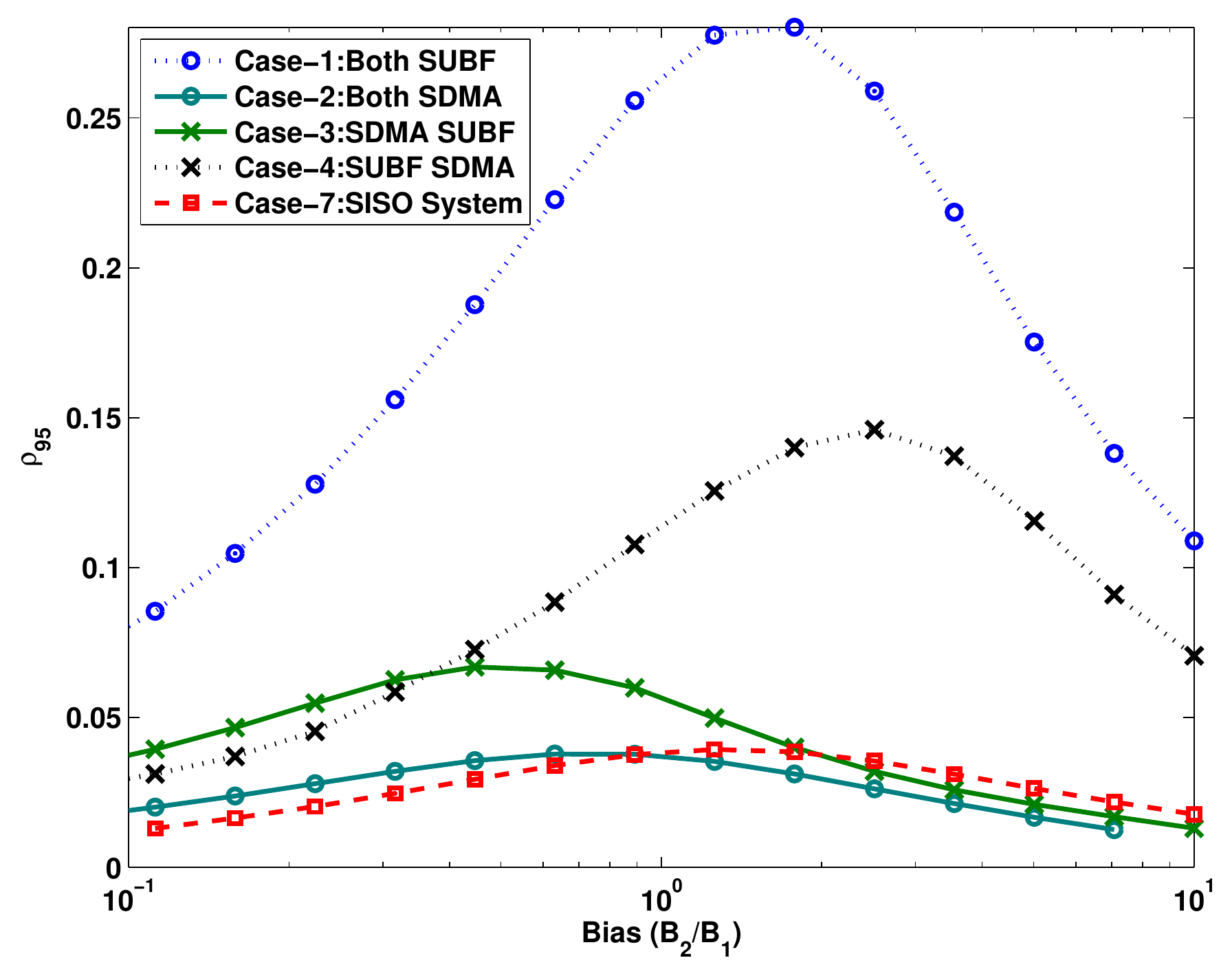}
\caption{The $5^{th}$ percentile rate $\rho_{95}$ versus bias $b=B_2/B_1$ for  a two-tier HetNet with $\alpha=4,  \lambda=[150,300],P_1=5P_2$.}
\label{fig:indi4}
\end{figure}

\subsection{Optimal Bias}
We now compute the optimal selection bias needed to maximize coverage probability. Recall that the BS selection based on the highest mean SINR may not always maximize coverage. We will also validate the selection bias approximations discussed in Section~\ref{sec:cell_selection} using these numerical results. Fig. \ref{fig:BiasOptThreeruns} presents the probability of coverage for target SIR $0$ dB as a function of $B_2/B_1$ for cases 1-4 corresponding to {\em 4-2 antenna configuration} and case 7 corresponding to {\em SISO configuration}. Fig. \ref{fig:cand} compares different selection (bias) candidate functions for all 7 simulation cases.
To compute the optimal bias, we simulate the system with each bias value between .01 to 100 and choose the bias which maximizes the probability of coverage. This optimal bias is then compared to the bias found from various candidate functions for different values of system parameters. It is evident that the candidate function \eqref{eq:biaspsibydelta} is a close match.

\subsection{Rate Coverage and Optimal Bias}

 We show the variation of rate coverage with bias in Fig. \ref{fig:BiasOptThreeruns_RC} for rate target 1 bps/Hz for {\em 4-2 antenna configuration}. It can be seen that rate coverage heavily depends on selection bias. Also, the rate coverage is maximized at a particular value of selection bias in multi-antenna HetNets, which is consistent with the intuition gained about SISO HetNets in~\cite{SinDhiJ2013}. It is worth noting that the optimal bias in this case is not necessarily the same as the one that maximizes coverage probability. We numerically compute these optimum biases for all 7 simulation cases and plot the resulting rate coverages in Fig. \ref{fig:BiasOptimized_ratecoverage}.  The SISO case results in the worst rate coverage. At the lower thresholds, case 1 (both tiers using SUBF) performs the best, while at higher thresholds, case 2 (both tiers using SDMA) is superior. Other two cases corresponding to the {\em 4-2 antenna configuration} perform in between these two extremes. For {\em 2-1 antenna configuration}, we observe a similar behavior where case 5 (first tier with SUBF) performs better for smaller rate thresholds while case 6 (first tier using SDMA) performs better for higher thresholds.

In Fig. \ref{fig:BiasOptimized_ratecoverage_comparison}, we compare rate coverage with optimum bias, the candidate bias function \eqref{eq:biaspsibydelta} and the worst bias (the bias  which gives the lowest rate coverage among the values between $[0.01,100]$) for simulation case 5. Recall that the case 5 corresponds to {\em 2-1 antenna configuration}, where first tier performs SUBF and the second performs SISO. Comparing the optimal rate coverage with the one achieved under worst bias again highlights the fact that choosing a poorly designed bias value can significantly degrade rate coverage. More interestingly, we observe that the candidate bias function that we derived for coverage maximization works reasonably well for this case as well. This can at least be used as a starting point for numerical search algorithms to find optimal bias. That being said, the difference in the two cases is higher at higher rate thresholds. This is explained in Fig.\ref{fig:BiasOptimized_ratecoverage_biasopt}, where we present the variation of optimal bias with rate threshold for all 7 simulation cases. Clearly, the optimal system design in this case depends upon the target rate, which in turn depends upon the target application, e.g., video. 

In Fig.\ref{fig:BiasOptimized_ratecoverage_biasopt}, we can also observe the effect of transmission schemes on optimal bias. Required selection bias values for moderate rate thresholds follow the following order: case $4>5>7>1>6>2>3$. This ordering is consistent with intuition. For instance, when the macrocell uses SUBF and femtocell uses SDMA or SISO, the coverage regions of macrocells are further expanded due to beamforming gain, which results in the need for higher selection bias for the second tier to balance load across the tiers compared to the case when the first tier performs SISO transmission. On the other hand, if we now assume that the first tier performs SDMA and the second tier performs SUBF, the coverage regions of the second tier are naturally expanded due to the beamforming gain, i.e., the load across the tiers is more balanced, which reduces the need for a high external bias for the second tier. In general, whenever small cells can use multi-antenna transmission for range expansion, e.g., by SUBF, the external bias required to balance load would be smaller compared to the SISO case. The other cases can also be explained similarly. 

\begin{figure}
\centering
\includegraphics[width=\columnwidth]{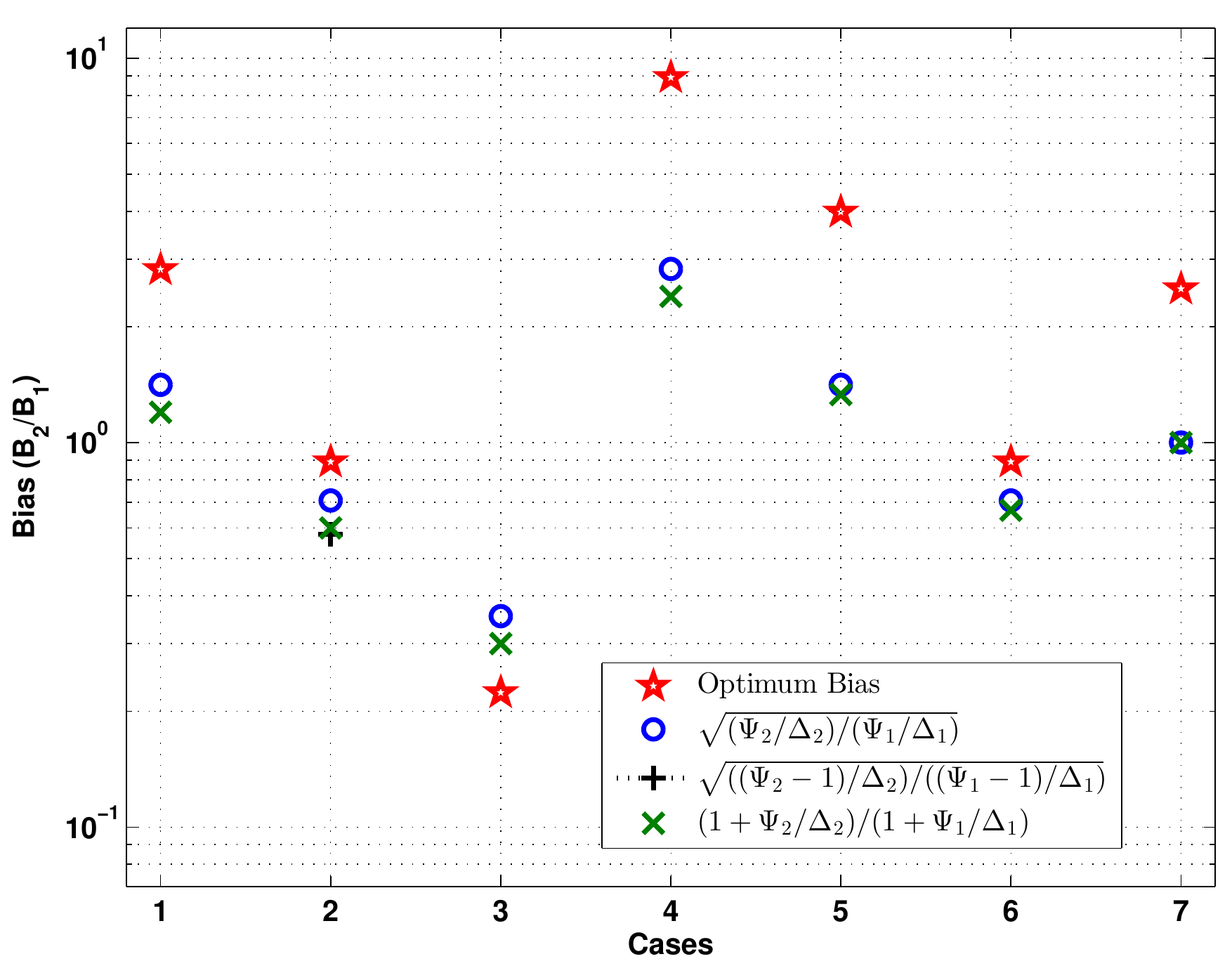}
\caption{Comparison of different candidates functions with numerically
optimized bias for target rate = 1bps/Hz,$\lambda = [150; 300]; P_1 = 5P_2$ for all 7 simulation cases.}
\label{fig:cand_rc}
\end{figure}

In Fig. \ref{fig:indi4}, we present the $5^{th}$ percentile rate, i.e., the target rate such that the $95\%$ users achieve rate higher than the target. The trends for the optimum bias are consistent with those discussed above. The rate coverage results are also consistent with those discussed for Fig. \ref{fig:BiasOptThreeruns_RC}. Finally, Fig. \ref{fig:cand_rc} compares the optimal bias for rate coverage and bias from the candidate functions for different cases. Recall that even though these candidate functions were derived for coverage maximization, they still provide good starting points for numerical search algorithms to find rate maximizing bias, as stated earlier in this section. 

\section{Conclusion}

In this paper, we have investigated downlink multi-antenna HetNets with flexible cell selection and shown that simple selection bias-based cell selection criterion closely approximates more complex selection rules to maximize mean SINR. Under this simpler cell selection rule, we derived exact expressions for coverage probability and rate achievable by a typical user. An approximation of the coverage optimal cell selection bias for each tier is also derived in closed form. Due to this connection with biasing, there is a natural expansion of coverage regions of small cells whenever small cells can use multi-antenna transmission for range expansion, e.g., by using beamforming. This leads to a natural balancing of load across tiers, which reduces the additional artificial cell selection bias needed to offload sufficient traffic to small cells. 


This work has numerous extensions. First, it is important to generalize this analysis to the case where mobile users have multiple antennas. This adds another dimension to system design and it is important to understand how to best use the additional degrees of freedom. Second, it is important to extend the analysis to include more general precoding techniques. Third, it is important to characterize the performance of multi-antenna HetNets under per BS power constraint. Recall that in this paper, we focused on the per user power constraint. Fourth, it is important to characterize the exact cell association regions and derive exact optimal bias expressions for important metrics, such as coverage and rate. Finally, it is important to extend these ideas to cellular uplink.

\appendix
\subsection{Proof for Lemma \ref{lemma:associationregion}: Association regions}\label{Proof:associationregion}
A user is associated with tier $k^{th}$ if other tier BS distances $d_j$ satisfies following relation for all $j$,
\begin{eqnarray*}
P_{rj}B_j&\le&P_{rk}B_k\\
P_j\Delta_j {\left(d_j\right)}^{-\alpha_j}B_j &\le& P_k \Delta_k {\left(d_k\right)}^{-\alpha_k}B_k\\
d_j &\ge&{\left(\widehat{P_j}\widehat{\Delta_j}\widehat{B_j}\right)}^{1/\alpha_j}{d_k}^{1/\widehat{\alpha_j}},
\end{eqnarray*}
where $\widehat{f_j}=f_j/f_k$.

\subsection{Derivation of Equation \eqref{eq:outageprob}}\label{Proofeqoutageprob}
For notational simplicity, let
\begin{align}
e_k(x_k,\phi\backslash x_k)&=\mathbbm{1}{(S_{x_k}\ge S_{y} \ \forall y\in\phi\backslash x_k)}\nonumber\\
&=\mathbbm{1}{\left(\|y\|\ge{(\widehat{P}_j\widehat{\Delta}_j\widehat{B}_j)}^{\frac{1}{\alpha_j}}\|x_k\|^{\frac{2}{\widehat{\alpha}_j}}\ \forall y\in \phi \backslash {x_k}\right)},\nonumber
\end{align}
then
\begin{align}
P_c&=\sum_{k=1}^{K}\mathbb{E}_{\Phi_k}\left[\mathbbm{1}\left(\frac{P_kh_{x_kk}}{\|x_k\|^{\alpha_k}}>T(N+I_{\Phi \backslash{x_k}})\right)e_k(x_k,\phi)\right]\nonumber\\
&\stackrel{(a)}{=}\sum_{k=1}^{K}\lambda_k\int_0^{\infty}\mathbb{P}\left[\frac{P_kh_{x_kk}}{\|x_k\|^{\alpha_k}}>T(N+I_{\Phi}),e_k(x_k,\phi)\right]{\rm d}x_k \nonumber\\
&=\sum_{k=1}^{K}\lambda_k\int_0^{\infty}\mathbb{P}\left(\frac{P_kh_{x_kk}}{\|x_k\|^{\alpha_k}}>T(N+I_{\Phi})|e_k(x_k,\phi)\right) \cdot\nonumber\\
&\ \ \ \ \  \mathbb{P}\left(\|y\|\ge{(\widehat{P}_j\widehat{\Delta}_j\widehat{B}_j)}^{\frac{1}{\alpha_j}}\|x_k\|^{\frac{1}{\widehat{\alpha}_j}}\ \forall y\in \Phi\right){\rm d}x_k\nonumber\\
&\stackrel{(b)}{=}\sum_{k=1}^{K}\lambda_k\int_0^{\infty}\mathbb{P}\left(P_kh_{x_kk}\|x_k\|^{-\alpha_k}>T(N+I_{\Phi^{'}})\right)\cdot\nonumber\\
& \ \ \ \ \  e^{-\sum_{j=1}^{K}\lambda_j\pi{(\widehat{P}_j\widehat{\Delta}_j\widehat{B}_j)}^{\frac{2}{\alpha_j}} \|x_k\|^{\frac{2}{\widehat{\alpha}_j}}}{\rm d}x_k\nonumber\\
&\stackrel{(c)}{=}2\pi\sum_{k=1}^{K}\lambda_k\int_0^{\infty}\mathbb{P}\left(P_kh_{xk}\|x\|^{-\alpha_k}>T(N+I_{\Phi^{'}})\right)\cdot\nonumber\\
&\ \ \ \ \  e^{-\sum_{j=1}^{K}\lambda_j\pi{(\widehat{P}_j\widehat{\Delta}_j\widehat{B}_j)}^{\frac{2}{\alpha_j}} x^{\frac{2}{\widehat{\alpha}_j}}}x{\rm d}x, 
\end{align}
where $\Phi^{'}$ is
\begin{equation}
\Phi^{'}=\cup_{k=1}^K{ \Phi_k \cap {B\left(0,{(\widehat{P}_j\widehat{\Delta}_j\widehat{B}_j)}^{\frac{1}{\alpha_j}} x^{\frac{1}{\widehat{\alpha}_j}}\right)}^c},
\end{equation}
where $B(0,r)$ is open ball around origin with radius $r$ and ${(.)}^c$ denotes the set complement.
Here $(a)$ is due to Cambell-Mecke's formula \cite{StoKenB1995} and Slivnyak's theorem, $(b)$ follows from the basic properties of a PPP, and $(c)$ by converting to polar coordinates:  $\|x_k\|\rightarrow x$. This completes the proof.

\subsection{Proof for Laplace Transform of the interference}\label{AppendixProofInter}
Laplace transform  of the sum interference caused by $j^{th}$ tier  BSs is due to all $j^{th}$ tier BSs outside the ball $B(0,r_j)$ where $r_j={\left(\widehat{P}_j\widehat{\Delta}_j\widehat{B}_j\right)}^{1/\alpha_j}{x}^{1/\widehat{\alpha_j}}$ is given by $\mathcal{L}_j(s)=\mathbb{E}[e^{-sI_j}]=$
\begin{align*}
&\mathbb{E}\left[\exp\left\{-s\sum_{y\in\Phi_j\backslash B(0,r_j)}P_{j}g_{yj}||y||^{-\alpha_j}\right\}\right]\\
=&\exp\left[-\lambda_j\int_{\mathbb{R}^2\backslash B(0,r_j)}{1-E_{g_{yj}}[e^{-sP_jg_{yj}||y||^{-\alpha_j}}]dy}\right]\\
=&\exp\left[-2\pi\lambda_j\int_{r_j}^{\infty}{1-E_{g_{rj}}[e^{-sP_jg_{rj}r^{-\alpha_j}}]rdr}\right]\\
=&\exp\left[-2\pi\lambda_j\int_{r_j}^{\infty}{\left(1-\frac{1}{{(1+sP_jr^{-\alpha_j})}^{\Psi_j}}\right)rdr}\right]\\
\stackrel{(a)}{=}&\exp\left[-2\pi{\lambda_j{(sP_j)}^{\frac{2}{\alpha_j}}\int_{t_j}^{\infty}{\left(1-\frac{1}{{(1+{t}^{-\alpha_j})}^{\Psi_j}}\right) t dt}}\right]\\
\stackrel{(b)}{=}&\exp\left[-2\pi\lambda_j{(sP_j)}^{\frac{2}{\alpha_j}}\sum_{m=1}^{\Psi_j} \binom{n}{k} \int_{t_j}^{\infty}{
    \left(
    \frac
       {{t}^{-\alpha_jm}}
        {{(1+{t}^{-\alpha_j})}^{\Psi_j}}
    \right) t dt}
\right]\\
\stackrel{(c)}{=}&\exp\left[-2\pi\frac{\lambda_j}{\alpha_j}{(sP_j)}^{\frac{2}{\alpha_j}}\sum_{m=1}^{\Psi_j} \binom{n}{k} \right. \cdot\\
& \left.\int_{u_j}^{1}{u^{\Psi_j-1-m-\frac{2}{\alpha_j}} {(1-u)}^{m+\frac{2}{\alpha_j}-1 } {\rm d}u}\right]\\
\stackrel{(d)}{=}&\exp\left[-2\pi\frac{\lambda_j}{\alpha_j}{(sP_j)}^{\frac{2}{\alpha_j}}\sum_{m=1}^{\Psi_j} \binom{n}{k}\right.\cdot\\ &\left.B'\left(\Psi_j+m-\frac{2}{\alpha_j},m+\frac{2}{\alpha_j},u_j\right)\right]
\end{align*}
with limits as
\begin{align*}
t_j&={(sP_j)}^{-1/\alpha_j}r_j\\
u_j&=\frac{1}{1+t_j^{-\alpha_j}}=\frac{1}{1+sP_j{\left(\widehat{P}_j\widehat{\Delta}_j\widehat{B}_j\right)}^{-1}{x}^{-\alpha_k}},
\end{align*}
where $(a)$ follows from substituting ${(sP_j)}^{-1/\alpha_j}r\rightarrow  t$, $(b)$ follows from binomial expansion and $(c)$ follows from $1/{(1+t^{-\alpha})}\rightarrow u$. In $(d)$, we defined $B'(a,b,c)$ as the complimentary incomplete Beta function as
\begin{align*}
B'(a,b,z)&=\int_{z}^{1}{u^{a-1}{(1-u)}^{b-1}{\rm d}u}.
\end{align*}
The Laplace transform of sum of aggregate interference from all BS and noise is equal to
\begin{align*}
\mathcal{L}_{IN}(s)&=E[e^{-s(I+N)}]=e^{-sN}\mathbb{E}\left[e^{-\sum_{j=1}^{K}{ I_j}}\right] \nonumber \\
&\stackrel{(e)}{=}e^{sN}\prod_{j=1}^{K} \mathbb{E}\left[e^{-{ I_j}}\right],
\end{align*}
where $(e)$ follows from independence of BS point processes  among tiers.

\subsection{Proof for derivatives of Laplace Transform of the interference plus noise}\label{AppendixProofInterDeriv}
The Laplace transform of noise plus interference   $\mathcal{L}_{IN}(s)$ can be written as $f(g(s))$ where $f(x)=\exp(x)$ and $g(s)=$
\begin{equation}
-sN+2\pi\sum_{j=1}^{K}{\lambda_j\int_{r_j}^{\infty}{\left(-1+\frac{1}{{(1+sP_jr^{-\alpha_j})}^{\Psi_j}}\right)r{\rm d}r}}\label{eq:a1}.
\end{equation}

Using Fa\`a di Bruno lemma, $n^{th}$ derivative can be written as $\frac{d^n\mathcal{L}_{IN}(s)}{ds^n}=$
\begin{equation}
\frac{{\rm d}^nf(g(s))}{{\rm d}s^n}=\sum_{\bar{m}\in M}{C(\bar{m})f^{1^T\bar{m}}(g(s))   \prod_{l=1}^{n}{{\left(g^{(l)}(s)\right)}^{m_l}}}.
\end{equation}
The $l^{th}$ derivatives of $g(s)$  can be computed as
\begin{eqnarray*}
g^{(l)}(s)&=&-N\mathbbm{1}_{l=1}+2\pi\sum_{j=1}^{K}\lambda_j{(-1)}^l\frac{(\Psi_j+l-1)!}{(\Psi_j-1)!}P_j^l
\\
&&\ \ \ \ \ \times\int_{r_j}^{\infty}{\frac{r^{1-l\alpha_j}}{{(1+sP_jr^{-\alpha_j})}^{\Psi_j+l}}{\rm d}r}\\
&\stackrel{(a)}{=}&-N\mathbbm{1}_{l=1}+2\pi\sum_{j=1}^{K}{{(-1)}^lD_j(l)P_j^l
{(sP_j)}^{\frac{2}{\alpha_j}-l}
}
\end{eqnarray*}
where $(a)$ can be computed using the similar transformations as used in computing Laplace transform of noise plus interference in Appendix \ref{AppendixProofInter}.

\subsection{Proof of Lemma \ref{thm:SINRdist}}\label{AppendixProofSINRDist}
Using expressions of $\mathcal{L}_{IN}$ and its derivatives, we can write \eqref{eqSINR1} as $P[\gamma_k(x) > T] =$
\begin{align*}
&\sum_{n=0}^{\Delta_k-1}{\frac{1}{n!}{(-s)}^n\frac{{\rm d}^n}{{\rm d}s^n}\mathbb{E}\left[e^{-s(I+N)}\right]}\\
&=\sum_{n=0}^{\Delta_k-1}\frac{1}{n!}{(-s)}^n
\mathcal{L}_{IN}(s)\sum_{\bar{m}\in M}{C(\bar{m})}\cdot\\
&{
\prod_{j=1}^{n}{{\left(N\mathbbm{1}_{l=1}+2\pi\sum_{j=1}^{K}{{(-1)}^l
D_j(l)
{(sP_j)}^{\frac{2}{\alpha_j}-l}
}\right)}^{m_j}}}
\end{align*}
with $s=TP_k^{-1}x^{\alpha_k}$.

\subsection{Proof for Coverage Probability: Interference limited case with $\alpha_j=\alpha\ \forall j$}\label{AppendixProofCov}
For this case, the coverage probability is
\begin{align}
P_c=&{2\pi \lambda_k}{}\int_{0}^{\infty}
\sum_{n=0}^{\Delta_k-1}\frac{1}{n!}{(TP_k^{-1})}^n\cdot\nonumber\\\nonumber
&
\sum_{\bar{m}\in M}C(\bar{m})F(\bar{m})e^{-\left[\sum_{j=1}^{K}{\lambda_j{(T\widehat{P_j})}^{\frac{2}{\alpha}}\calCfunc_j}\right]x^2}x^{2\sum{m_l}}\cdot \\\nonumber
& xe^{
-\pi \sum_{j=1}^{K}{\lambda_j{(\widehat{P_j}\widehat{\Delta_j}\widehat{B_j})}^{2/\alpha}x^{2}}
}{\rm d}x\\\nonumber
=&{2\pi \lambda_k}
\sum_{n=0}^{\Delta_k-1}\frac{1}{n!}{(TP_k^{-1})}^n\\\nonumber
&\sum_{\bar{m}\in M}{C(\bar{m})F(\bar{m})}\int_{0}^{\infty}
e^{-\left[\sum_{j=1}^{K}
{\lambda_j{(T\widehat{P_j})}^{\frac{2}{\alpha}}\calCfunc_j}\right]
x^2}\cdot\\\nonumber
& e^{ -\left[\pi \sum_{j=1}^{K}{\lambda_j{(\widehat{P_j}\widehat{\Delta_j}\widehat{B_j})}^{2/\alpha}}\right]x^2}
 x^{1+2\sum{m_l}}{\rm d}x\\\nonumber
\stackrel{(a)}{=}&\Scale[0.95]{{\pi \lambda_k}{}
\sum_{n=0}^{\Delta_k-1}\frac{1}{n!}{(TP_k^{-1})}^n}\cdot\\\nonumber
&\Scale[0.95]{\sum_{\bar{m}\in M}{ \frac{C(\bar{m})F(\bar{m})\Gamma(\sum{m_l}+1)}
{\left[\sum_{j=1}^{K}{\lambda_j{(T\widehat{P_j})}^{\frac{2}{\alpha}}\calCfunc_j}+\pi \sum_{j=1}^{K}{\lambda_j{(\widehat{P_j}\widehat{\Delta_j}\widehat{B_j})}^{2/\alpha}}\right]^{\sum{m_l}+1}}}}\\\nonumber
\end{align}
where $(a)$ follows from $\int_0^{\infty}{e^{-ax^2}x^{2n+1}}=1/{(a)}^{n+1}$.

\bibliographystyle{IEEEtran}	
\bibliography{Gupta,Dhillon,MIMO}

\begin{IEEEbiography} [{\includegraphics[width=1in,height=1.25in,clip,keepaspectratio]{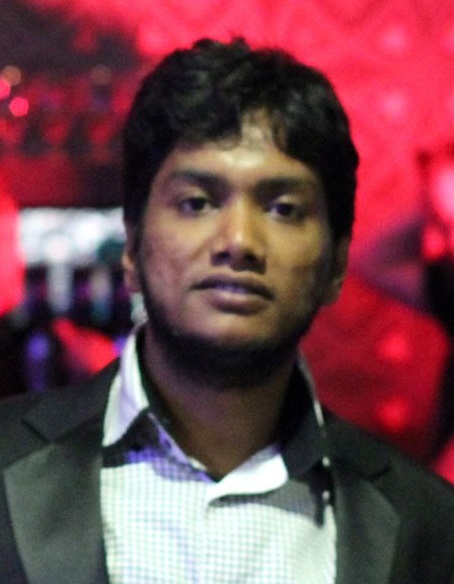}}]
{Abhishek K. Gupta} 
received the B.Tech. and M.Tech. degrees in Electrical Engineering from IIT Kanpur, India, in 2010. He is currently a Ph.D. student at The University of Texas at Austin, where his research has focused on stochastic geometry and its applications in wireless communication. His other research interests include multiuser MIMO systems and optimization. He was recipient of GE-FS Leadership Award by General Electric (GE) Foundation and Institute of International Education in 2009. He has held summer internships at Huawei technologies, NJ, and Cranfield University, UK. He is author of the book {\em MATLAB by Examples} (2010).
\end{IEEEbiography}

\begin{IEEEbiography} [{\includegraphics[width=1in,height=1.25in,clip,keepaspectratio]{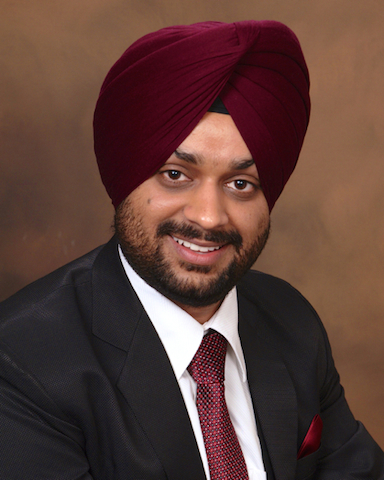}}]
{Harpreet S. Dhillon} 
(S'11, M'13) is an Assistant Professor of Electrical and Computer Engineering at Virginia Tech, Blacksburg, VA. He received the B.Tech. degree in Electronics and Communication Engineering from Indian Institute of Technology (IIT) Guwahati, India, in 2008; the M.S. degree in Electrical Engineering from Virginia Tech, Blacksburg, VA, USA, in 2010; and the Ph.D. degree in Electrical Engineering from the University of Texas (UT) at Austin, TX, USA, in 2013. After a postdoctoral year at the University of Southern California (USC), Los Angeles, CA, USA, he joined Virginia Tech in August 2014. He has held summer internships at Alcatel-Lucent Bell Labs in Crawford Hill, NJ, USA; Samsung Research America in Richardson, TX, USA; Qualcomm Inc. in San Diego, CA, USA; and Cercom, Politecnico di Torino in Italy.

Dr. Dhillon has been a co-author of three best paper award recipients: the 2014 IEEE Leonard G. Abraham Prize, the 2014 European Wireless Best Student Paper Award, and the 2013 IEEE International Conference in Communications (ICC) Best Paper Award in Wireless Communications Symposium. 
He is a recipient of the USC Viterbi Postdoctoral Fellowship, the 2013 UT Austin Wireless Networking and Communications Group (WNCG) leadership award, the UT Austin Microelectronics and Computer Development (MCD) Fellowship, and the 2008 Agilent Engineering and Technology Award, a national award for the best undergraduate research thesis in India. His research interests include communication theory, stochastic geometry, and wireless ad hoc and heterogeneous cellular networks.
\end{IEEEbiography}

\begin{IEEEbiography} [{\includegraphics[width=1in,height=1.25in,clip,keepaspectratio]{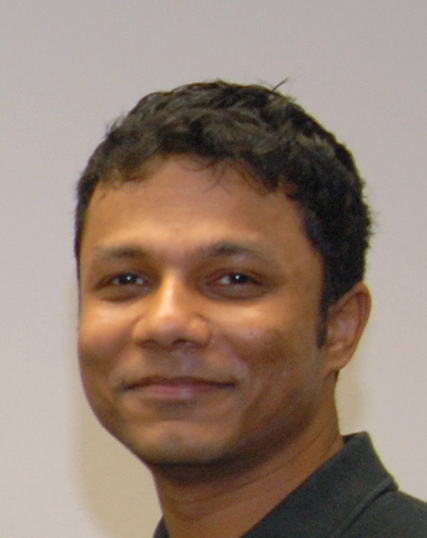}}]
{Sriram Vishwanath} 
 received his B. Tech. degree in Electrical Engineering from the Indian Institute of Technology (IIT), Madras, India in 1998, the M.S. degree in Electrical Engineering from California Institute of Technology (Caltech), Pasadena USA in 1999, and the Ph.D. degree in Electrical Engineering from Stanford University, Stanford, CA USA in 2003. He is currently a Professor in the Department of Electrical and Computer Engineering at The University of Texas, Austin, USA. His research interests include network information theory, wireless systems and mobile systems.  He has served on the board of directors and/or co-founded several startups, including M87 Inc., Lynx Laboratories Inc., Azul Mobile Inc., Agilux Systems Inc., Accordion Health Inc. and Sunfish Inc.; and was the 2014 University of Texas Entrepreneur of the Year.

Sriram received the NSF CAREER award in 2005 and the ARO Young Investigator Award in 2008. He is the co-recipient of the 2005 IEEE Joint Information Theory Society and Communications Society best paper award. He has served as the general chair of IEEE Wireless Network Coding conference (WiNC) in 2010, the general co-chair of the IEEE Information Theory School in 2011, the local arrangements chair of ISIT 2010 and the guest editor-in-chief of Transactions on Information Theory special issue on interference networks.
\end{IEEEbiography}

\begin{IEEEbiography} [{\includegraphics[width=1in,height=1.25in,clip,keepaspectratio]{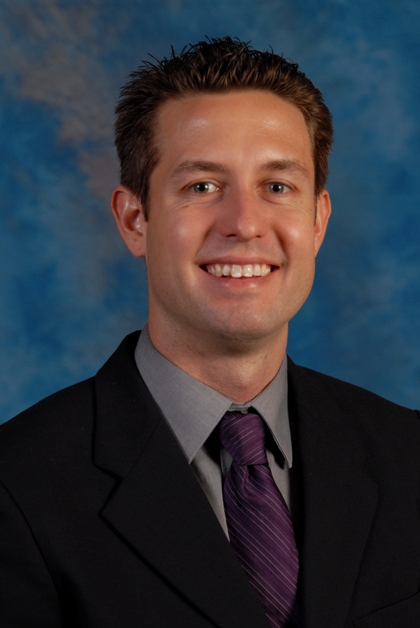}}]
{Jeffrey Andrews} 
 (S'98, M'02, SM'06, F'13) received the B.S. in Engineering with High Distinction from Harvey Mudd College, and the M.S. and Ph.D. in Electrical Engineering from Stanford University.  He is the Cullen Trust Endowed Professor (\#1) of ECE at the University of Texas at Austin, Editor-in-Chief of the IEEE Transactions on Wireless Communications, and Technical Program Co-Chair of IEEE Globecom 2014. He developed Code Division Multiple Access systems at Qualcomm from 1995-97, and has consulted for entities including Verizon, the WiMAX Forum, Intel, Microsoft, Apple, Samsung, Clearwire, Sprint, and NASA.  He is a member of the Technical Advisory Boards of Accelera and Fastback Networks, and co-author of the books {\em Fundamentals of WiMAX} (Prentice-Hall, 2007) and {\em Fundamentals of LTE} (Prentice-Hall, 2010).  

Dr. Andrews received the National Science Foundation CAREER award in 2007 and has been co-author of ten best paper award recipients: ICC 2013, Globecom 2006 and 2009, Asilomar 2008, European Wireless 2014, the 2010 IEEE Communications Society Best Tutorial Paper Award, the 2011 IEEE Heinrich Hertz Prize, the 2014 EURASIP Best Paper Award, the 2014 IEEE Stephen O. Rice Prize, and the 2014 IEEE Leonard G. Abraham Prize.  He is an IEEE Fellow and an elected member of the Board of Governors of the IEEE Information Theory Society. 
\end{IEEEbiography}
\end{document}